\documentclass[conference]{IEEEtran}
\IEEEoverridecommandlockouts
\usepackage{cite}
\usepackage{amsmath,amssymb,amsfonts}
\usepackage{algorithmic}
\usepackage{graphicx}
\usepackage{textcomp}
\usepackage{xcolor}
\usepackage{booktabs}
\usepackage{multirow}
\usepackage{float}
\usepackage{algorithm}
\usepackage{url}
\usepackage{caption}
\usepackage{tikz}
\usetikzlibrary{positioning,shapes,arrows.meta,calc}

\def\BibTeX{{\rm B\kern-.05em{\sc i\kern-.025em b}\kern-.08em
    T\kern-.1667em\lower.7ex\hbox{E}\kern-.125emX}}

\begin{document}

\title{Physical Side-Channel Attacks on Edge LLMs: Extracting Output Tokens and User Prompts}
\author{
\IEEEauthorblockN{Najeh Alhalawani}
\IEEEauthorblockA{\textit{Radboud University}}
\and
\IEEEauthorblockN{Gunes Acar}
\IEEEauthorblockA{\textit{Radboud University}}
\and
\IEEEauthorblockN{Zhuoran Liu}
\IEEEauthorblockA{\textit{University of Amsterdam }}
\and
\IEEEauthorblockN{Lejla Batina}
\IEEEauthorblockA{\textit{Radboud University}}
}

\maketitle

\begin{abstract}
The deployment of large language models (LLMs) on edge devices is largely motivated by privacy concerns over sharing sensitive user data with cloud servers. 
In this paper, we demonstrate that keeping inference local does not eliminate privacy risks: an adversary with physical access to the device may exploit hardware leakage to recover sensitive information without compromising or modifying the software system.
In particular, we present an electromagnetic (EM) side-channel attack targeting LLM inference on edge devices. 
Targeting LLaMA-3.2-1B-Instruct running on a Raspberry Pi 5, we demonstrate that EM traces collected during inference can leak both \emph{generated tokens} and \emph{user prompts}. 
We propose three attacks under different threat assumptions.
First, a \emph{profiling attack} that identifies generated tokens among 50 medical and privacy-related keywords with 97.2\% accuracy on profiled words. 
Second, a \emph{reconstruction attack} recovers the model's hidden representations and uses them to predict tokens from the full 128k-token vocabulary, achieving 94.3\% accuracy on profiled classes. 
For unseen words, the correct token is ranked within the first 100 positions of the 128k-token vocabulary in 67\% of cases, with a median rank of 54. 
Finally, we present a \emph{hidden-state inversion attack} that uses the recovered hidden state to reconstruct user prompts semantically. 
Our proposed attacks require only passive EM side-channel measurements and a limited number of traces during model training. 
These results demonstrate that local execution alone may not guarantee the confidentiality of generated tokens and user inputs. 

\end{abstract}


\section{Introduction}
\label{ch:introduction}

Large language models (LLMs) are no longer confined to cloud data centers. Increasing concerns about data privacy, latency, and intellectual property have motivated the deployment of LLM inference directly on edge devices. Local execution keeps sensitive information on the device, reduces reliance on external services, and enables applications requiring low latency or offline operation. 
At the same time, advances in model compression, quantization, and efficient inference libraries have made it increasingly feasible to run capable language models directly on edge hardware~\cite{gerganov2023llamacpp, shen2025agilequantactivationguidedquantizationfaster}. Systems such as the Raspberry Pi~5~\cite{raspberrypi5}, NVIDIA Jetson Orin Nano~\cite{nvidia_jetson_orin}, and dedicated neural processing accelerators such as the Hailo AI~\cite{hailo8l_aikit} can support local LLM inference. 
Edge-deployed LLMs are increasingly adopted for applications handling sensitive data, such as on-device medical diagnostics and health monitoring~\cite{prajapati2024medaide, su2024physiollm, zhong2025considerations}, offline legal analyzers~\cite{demir2025legalguardian}, and confidential mental health companions~\cite{xue2026mindchat, mandal2025towards}. In these use cases, local inference is often expected to improve privacy by keeping user prompts and generated outputs on-device rather than sending them to remote servers. However, local execution does not eliminate physical side-channel vulnerabilities that can leak sensitive data through power consumption or electromagnetic emanations.

In this paper, we demonstrate that an adversary with only passive physical proximity to the device can exploit physical side channels, such as electromagnetic emanations generated during model execution, to recover sensitive information processed by the LLM. 

All computing devices exhibit physical side effects during operation. As processors execute instructions, fluctuations in power draw and logic switching produce measurable electromagnetic emanations~\cite{agrawal2002em, PEETERS200752}. Electromagnetic side-channel analysis uses these emanations to infer information about the underlying computation. Previous work has shown that EM measurements can reveal sensitive values from cryptographic implementations, including secret keys~\cite{10.1007/3-540-45418-7_17, 10.1007/3-540-44709-1_21, agrawal2002em, 9586131, wang2024amplitude}. Similar leakage has also been observed in machine-learning computations, which allow attackers to extract the model's architecture, weight parameters, or user inputs~\cite{batina2018csinn, 9615240, horvath2025barracuda, horvath2025kraken, liu2025scaar}. However, it remains unclear whether EM leakage from an LLM running on an edge device can reveal the tokens generated during inference or information about the user's prompt. We therefore examine whether physical emanations from an edge device's CPU leak information about the LLM's outputs and its internal hidden representations during inference.

The model's final hidden representation is used by the language-model (LM) head projection layer to compute scores for the next token across the vocabulary. 
Since hidden representations are computed from preceding input sequences, they contain information derived from the user's prompt. This raises the possibility that physical leakage can reveal information about the LLM's input, in addition to the generated output. Building on this, we demonstrate that the recovered representations can be leveraged to infer the semantic content of user prompts.

We study LLaMA-3.2-1B-Instruct~\cite{dubey2024llama3} running on a Raspberry Pi 5~\cite{raspberrypi5} using electromagnetic side-channel analysis. We trace the \texttt{llama.cpp}~\cite{gerganov2023llamacpp} inference pipeline to identify the computational stages associated with token-dependent leakage and apply deep learning to recover information from the measured traces. We consider both direct output token classification and reconstruction of the hidden representation, which subsequently enables prompt recovery directly from the inferred LLM hidden states.


Our study makes three principal contributions:

\begin{enumerate}
  \item \textbf{Characterization of EM leakage during edge LLM inference.} \\
  We instrument the \texttt{llama.cpp} inference pipeline to trace the execution flow and isolate the electromagnetic leakage produced by distinct computational stages. Our analysis identifies the LM Head projection as a source of token-dependent leakage on our target hardware.

  \item \textbf{Deep-learning-based recovery of generated tokens from CPU emanations.} \\
    We develop two token-recovery approaches with different prediction spaces. The first classifies traces using a fixed set of profiled (target) tokens. The second reconstructs the hidden representation used by the LM Head and uses it to predict tokens over the model’s full vocabulary, including tokens not present during profiling.  
  \item \textbf{Recovery of semantic information about user prompts from physical leakage.} \\
    We investigate whether the recovered hidden representation can be inverted to infer information about the user prompt. Training an inversion model, we recover semantic information about user prompts from the representation reconstructed from the physical side channel.
\end{enumerate}

\section{Background}
\label{sec:background}

\subsection{Autoregressive LLM Inference}

Large language models (LLMs) generate text autoregressively by predicting the next token from the preceding context~\cite{dubey2024llama3}. During inference, the execution splits into two distinct phases: \emph{prefill} and \emph{decode}. In the prefill phase, the model processes the entire input prompt at once to establish the initial key-value (KV) cache and compute the first token. In the subsequent decode phase, the model generates tokens sequentially. For each new token, the model executes a forward pass that culminates in a final hidden state. 
Therefore, the execution follows a sequential pipeline: the context yields a hidden state, the state predicts the next token, and the system appends the new token back into the context. This repetitive decode phase presents a prominent side-channel target. Since the model executes an isolated, token-specific computation at every step, it causes distinct physical emanations. By recording these emanations, an adversary can extract the internal state determining the model's immediate output.

\subsection{Attack Formulations}
The LM Head component of LLMs connects the final hidden state to the next-token prediction. At decoding step $t$, the transformer layers produce a final hidden state $h_t \in \mathbb{R}^{d}$, where $d$ is the hidden dimension of the model. This contextual representation depends on the preceding sequence and is utilized to predict the next token. Specifically, the model passes $h_t$ through the LM Head, which performs a matrix-vector multiplication against the vocabulary matrix $W \in \mathbb{R}^{|\mathcal{V}| \times d}$, where $|\mathcal{V}|$ is the size of the vocabulary, to yield a vector of logits $z_t$:
\begin{equation}
z_t = W h_t.
\label{eq:lmhead-bg}
\end{equation}
The inference runtime subsequently converts these logits into a probability distribution over the vocabulary from which the next token is sampled:
\begin{equation}
h_t \rightarrow W h_t \rightarrow z_t \rightarrow p(y_t|y_{< t}) \rightarrow y_t.
\end{equation}

For LLaMA-3.2-1B-Instruct, the LM Head transforms a 2048-dimensional representation into scores across a $128\,256$-token vocabulary. The structural bottleneck presented by $h_t$ enables two distinct side-channel attack formulations to recover the generated token $y_t$ from an electromagnetic trace $x_{\mathrm{EM}}$. 

The first formulation is a \textbf{closed-set classification} attack:
\begin{equation}
x_{\mathrm{EM}} \rightarrow y_t,
\end{equation}
which maps the trace directly to a token $y_t$ belonging to a predefined set of $C$ profiled (target) classes (e.g., $C=50$). Although straightforward, this approach restricts recovery to words that have been profiled in advance.

The second formulation is a \textbf{representation recovery} attack:
\begin{equation}
x_{\mathrm{EM}} \rightarrow \hat{h}_t \rightarrow W \hat{h}_t \rightarrow \hat{y}_t.
\end{equation}
In this approach, the continuous 2048-dimensional hidden state $\hat{h}_t$ is estimated directly from the physical leakage. By projecting this estimated state through the model's public vocabulary matrix $W$, a complete logit vector is reconstructed to recover the token $\hat{y}_t$. This formulation removes the fixed-class bottleneck and enables token ranking over the model's full vocabulary.

\subsection{Physical Side-Channel Leakage}

The LM Head projection $W h_t$ constitutes a viable target for electromagnetic side-channel analysis for three reasons. First, it is a large matrix-vector multiplication that dominates the final stages of the decode step. Second, this mathematical operation is executed deterministically at every decoding step. Third, it utilizes quantized integer operations on fixed memory layouts. These factors create a prolonged, consistent, and data-dependent computation window. Consequently, the physical emanations generated during this phase correlate with the specific values in $h_t$ and can be passively acquired as a time-domain trace using a near-field EM probe positioned above the processor package.

\section{Related Work}
\label{ch:related_work}

\vspace{0.5ex}\noindent\textbf{Side-Channel Attacks on Machine Learning Models.}
Several lines of work target machine learning models themselves as the sensitive asset, seeking to recover model architecture, weight parameters, or user inputs from side-channel observations. Batina et al.~\cite{batina2018csinn} introduced the CSI~NN framework, demonstrating that the architecture of a neural network running on an ARM Cortex-M microcontroller can be reverse-engineered from EM and power traces. A recent comprehensive survey categorizes this literature, detailing architecture extraction, parameter extraction, and input recovery attacks across microcontrollers, CPUs, FPGAs, and GPUs~\cite{horvath2024sok}.

Horvath et al.~\cite{horvath2025barracuda} extended weight extraction to GPU hardware, introducing the BarraCUDA attack on the NVIDIA Jetson Nano using correlation EM analysis to recover convolutional network weights. The Kraken work by Horvath et al.~\cite{horvath2025kraken} pushed this further to specialized Tensor Core units, showing weight extraction from DNNs via higher-order EM correlation attacks. Liu et al.~\cite{liu2025scaar} introduced ScaAR, a generic physical side-channel attack on edge neural network implementations that extracts user inputs via EM emanations, directly targeting the privacy of inputs rather than model intellectual property (IP). Wang et al.~\cite{wang2024imc} extended the attack surface to in-memory computing architectures, showing that fixed weight-memory access patterns in RRAM IMC systems leak layer type and architecture information.

These works establish that physical side channels can reveal information about machine learning architectures and their inputs. \emph{However, they do not target the information produced by generative models during inference.}

\vspace{0.5ex}\noindent\textbf{Side-Channel Attacks on LLM Deployments.}
Recent work addresses side-channel attacks on LLM deployments, but through microarchitectural and network channels rather than physical emanations. Gao et al.~\cite{gao2025iknow} exploited CPU cache access patterns during the token embedding operation to infer generated token values, while Adiletta and Sunar~\cite{adiletta2025spill} used Flush+Reload to monitor cache activity shared with the victim and leak embedding-vector accesses, achieving up to 40\% recovery for English text. McDonald and Bar Or~\cite{mcdonald2025whisper} introduced Whisper Leak, a network-level side-channel that infers prompt topics from packet-size and timing patterns in encrypted streaming LLM traffic. These approaches rely on monitoring shared resources or network-level measurements. \emph{Our work, in contrast, demonstrates how token leakage can be exploited through purely physical electromagnetic emanations for open-vocabulary scenarios, without requiring any interaction with the device's software interfaces during the attack phase.}

\vspace{0.5ex}\noindent\textbf{Reconstructing Text from Model Internals.}
A separate line of research investigates whether text can be recovered from intermediate representations produced by language models. Morris et al.~\cite{morris2023vec2text} showed that sentence embeddings can be inverted into text using a sequence-to-sequence model. Later work extended this idea to language models by reconstructing prompts from next-token probability distributions~\cite{morris2024languagemodelinversion}. These approaches assume direct access to internal model representations or outputs through software interfaces. \emph{In contrast, we investigate whether equivalent information can be recovered from physical electromagnetic leakage, where the target representation must first be estimated from side-channel measurements rather than directly obtained from the model.}



\label{sec:research-gap}
\vspace{0.5ex}\noindent\textbf{Research Gap.}
Prior work on side-channel attacks against machine learning primarily recovers a model's architecture, weight parameters, or user inputs. They do not target the intermediate representations or tokens produced by LLMs during generation. While recent microarchitectural attacks such as cache monitoring~\cite{gao2025iknow, adiletta2025spill} can recover generated LLM tokens, they require access to shared hardware resources, and network-based attacks require the ability to observe the victim's network traffic. These threat vectors are inapplicable to an adversary with only passive physical proximity to an offline edge device. 

To our knowledge, no prior work has demonstrated that the tokens generated by an on-device LLM can be recovered purely from raw physical electromagnetic emanations. We fill this gap by targeting the LM Head projection layer rather than cache lookups. Our work demonstrates that this physical leakage vector is both practical and effective, requiring only close physical proximity and the ability to capture the device's emanations.

\begin{figure}
  \centering
  \includegraphics[width=\linewidth]{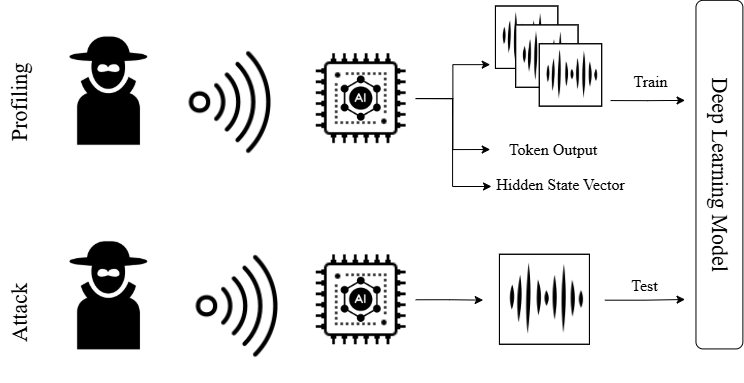}
  \caption{Overview of the profiling side-channel attack. During profiling (top), the adversary collects EM traces from a reference device, labeled with the corresponding output token and hidden state, to train a classification or generative model. During the attack (bottom), the same probe is applied to the victim device, and an unlabeled trace is analyzed to infer the generated token.}
  \label{fig:profiling-attack}
\end{figure}
\begin{figure*}
  \centering
  \includegraphics[width=0.8\linewidth]{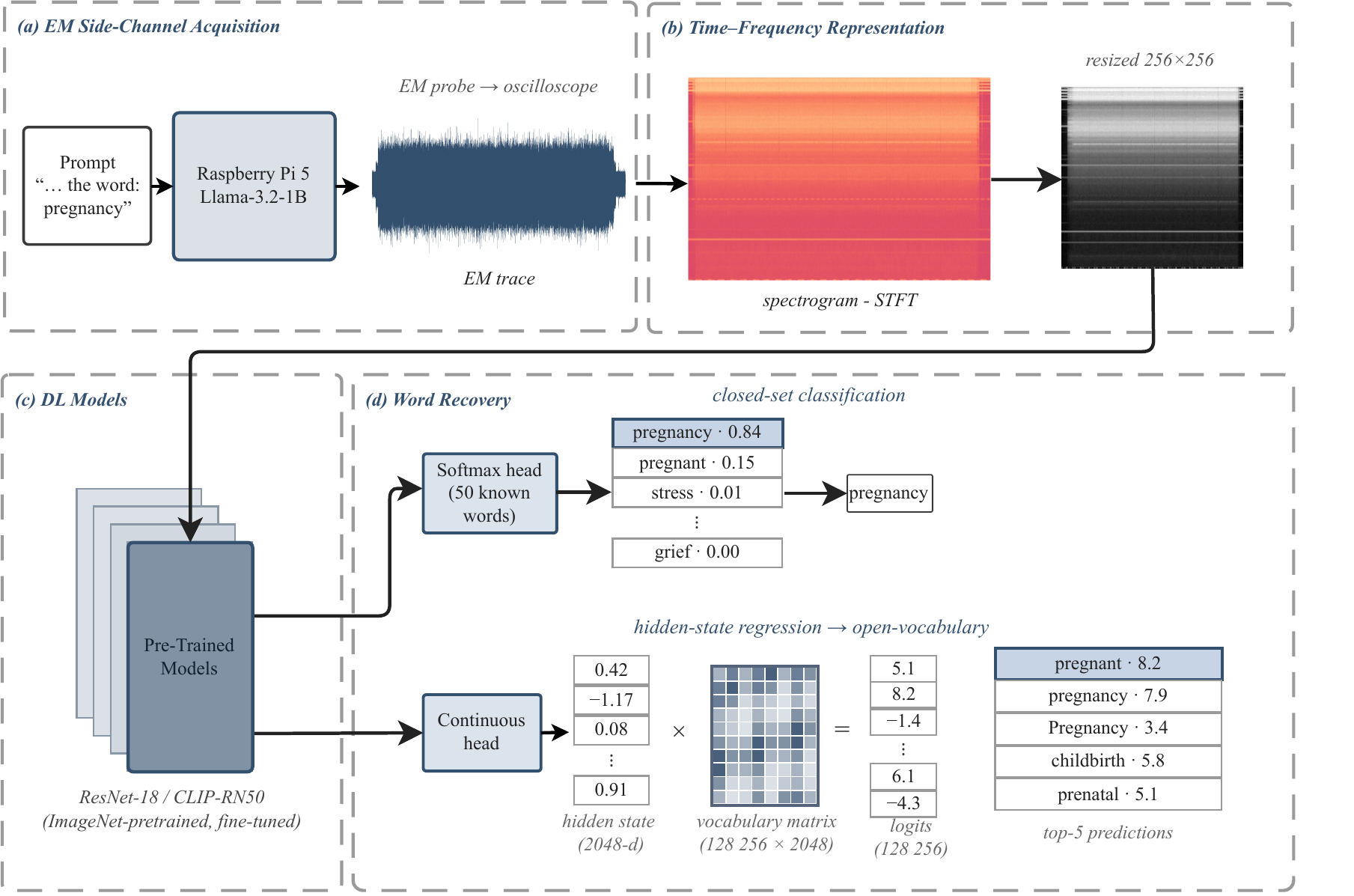}
  \caption{Overview of the methodology pipeline. \textbf{(a)} EM traces are captured during LLM inference on an edge device. \textbf{(b)} Raw 1D traces are transformed into 2D STFT spectrograms. \textbf{(c)} Pre-trained vision models extract features from these spectrograms. \textbf{(d)} Tokens are recovered using either closed-set classification or open-vocabulary prediction by generating the LLM's internal hidden states and multiplying by the vocabulary matrix to produce the final logits.}
  \label{fig:dlsca-framework-pipeline} 
\end{figure*}

\section{Threat Model}
\label{ch:threat-model}

This work assumes a physical, passive adversary. 
The adversary places a near-field electromagnetic probe near the target device and records EM traces during LLM inference. 
During the attack phase, the adversary has no access to the victim device's software, operating system, memory contents or network communication. 
The attack requires no hardware modification or software exploitation.
Trace acquisition happens while the device operates normally.

Local execution is often chosen to keep user data private. However, this privacy is threatened if an adversary can recover tokens directly from physical emanations. Unlike strictly controlled server environments, edge devices are frequently deployed in public, accessible locations and are typically less like to have robust physical protections. Consequently, an adversary only needs physical proximity to record electromagnetic emanations and completely bypass logical software security measures~\cite{zhong2025considerations,10.1145/3786170.3788387}.

Furthermore, this work distinguishes between a closed-set classification attack and a full-vocabulary representation recovery attack. While standard profiling attacks merely match traces against explicitly predefined classes, our threat model for full-vocabulary token recovery represents a stronger capability. In this formulation, the adversary uses the profiled traces merely as a corpus to construct a continuous mapping of the hidden-state representation space. This maps the physical leakage to the conceptual space, opening the door to an open-world attack that allows the adversary to recover arbitrary tokens across the entire vocabulary of the LLM, without having profiled the specific generated token beforehand.

\vspace{0.5ex}\noindent\textbf{Profiling Side-Channel Attack.}
The attack follows a standard profiling side-channel paradigm~\cite{chari2002template}, as shown in Figure~\ref{fig:profiling-attack}. During profiling, the adversary uses a reference device to collect labeled EM traces and the LLM's final hidden representation to train a model offline. In the attack phase, the adversary targets the victim device, requiring only physical proximity to acquire EM traces without interacting with the victim's software.

\usetikzlibrary{calc}

\section{Methodology}
\label{ch:model}

This section details the experimental platform used to capture electromagnetic traces and the methodology to turn those traces into a working attack. It describes the hardware and software setup, validates that the EM traces leak information, builds the dataset and feature representations, trains the attack models, and analyzes where the models locate leakage. Figure~\ref{fig:dlsca-framework-pipeline} provides an overview of this framework pipeline.

\label{sec:platform}

The target device is a Raspberry Pi~5 with a Cortex-A76 CPU~\cite{raspberrypi5}. The target model is LLaMA~3.2~1B Instruct~\cite{dubey2024llama3}, using Q4\_K\_M quantization and executed via \texttt{llama.cpp}~\cite{gerganov2023llamacpp}. We use a custom inference binary that calls the \texttt{llama.h} API to insert hardware triggers during execution. The LM Head utilizes Q6\_K weights and Q8\_K activations during its matrix-vector multiplication, executing via optimized ARM NEON dot-product instructions~\cite{arm_neon_optimization}.

\subsection{Dataset Construction}
\label{sec:datasets}

To systematically evaluate the side-channel leakage, we construct two distinct datasets. Every trace is captured while the model generates a known keyword. The prompt is constructed so that the reply is exactly two tokens: a fixed lead-in word followed by the keyword. The keyword is therefore always the second generated token, which fixes the position of the computation and makes traces directly comparable.

\vspace{0.5ex}\noindent\textbf{Primary 50-Class Dataset.}
The primary corpus targets closed-set token classification and hidden-state recovery across a sensitive vocabulary. We collected $120\,750$ traces across 50 health- and privacy-sensitive keywords (e.g., \texttt{cancer}, \texttt{HIV}, \texttt{depression}), yielding $2\,415$ traces per class. These traces are collected using a single fixed prompt template:
\label{quote:template}
\begin{quote}
\texttt{Say the words `hello \{W\}' and nothing else, with no punctuation.}
\end{quote}
where \texttt{\{W\}} is the keyword. The traces are cropped based on the hardware General Purpose Input/Output (GPIO) trigger to isolate the LM Head window, producing arrays of length $T\approx11.25\times10^6$ samples. A 70/15/15 split allocates $84\,524$ traces to training and $18\,113$ each to validation and test. Note that the GPIO trigger is used only for experimental alignment during trace collection and is not assumed to be available to the adversary during the attack phase.

\vspace{0.5ex}\noindent\textbf{Context-Independence Dataset (17 Templates).}
A secondary corpus is constructed specifically to prove that the hidden state leakage is independent of the preceding prompt context (evaluated in Section~\ref{sec:hidden-consistency}). This dataset varies the instruction across 17 distinct templates while keeping the keyword in the same generation position. The templates differ in both the instruction phrasing and the lead-in word, forcing the model to generate the same keyword from different preceding contexts. This ensures the recovery model learns the representation of the token itself, rather than overfitting to a specific prompt structure.
\label{sec:vocabulary}




\label{sec:measurement}

\begin{figure}
  \centering
  \includegraphics[width=0.8\linewidth]{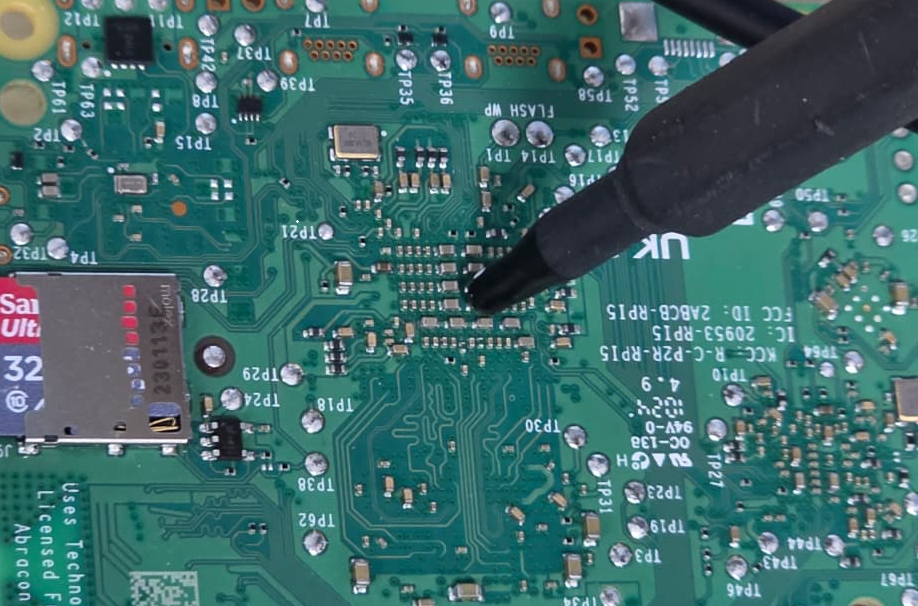}
  \caption{Near-field H-field loop probe positioned on the underside of the Raspberry Pi~5 PCB directly above the BCM2712 processor die.}
  \label{fig:setup-board}
\end{figure}

To produce a stable LM Head computation window, we fix the Cortex-A76 CPU frequency at 1500~MHz. At this clock, the matrix-vector multiplication in the LM Head takes approximately 22.5~ms, which increases the number of samples available during the computation window. We capture EM traces using a LeCroy WaveRunner~8404M-MS oscilloscope~\cite{lecroy_waverunner8404m}, a 4-channel, 4~GHz bandwidth instrument. We configure the measurement channel C2 at 28~mV/division with a 50~$\Omega$ input impedance. We position a passive near-field magnetic (H-field) loop probe on the underside of the Raspberry Pi~5 PCB, directly above the BCM2712 processor die, targeting the region where we observed the strongest EM activity (Figure~\ref{fig:setup-board}).

\begin{figure}[htbp]
  \centering
  \includegraphics[width=\linewidth]{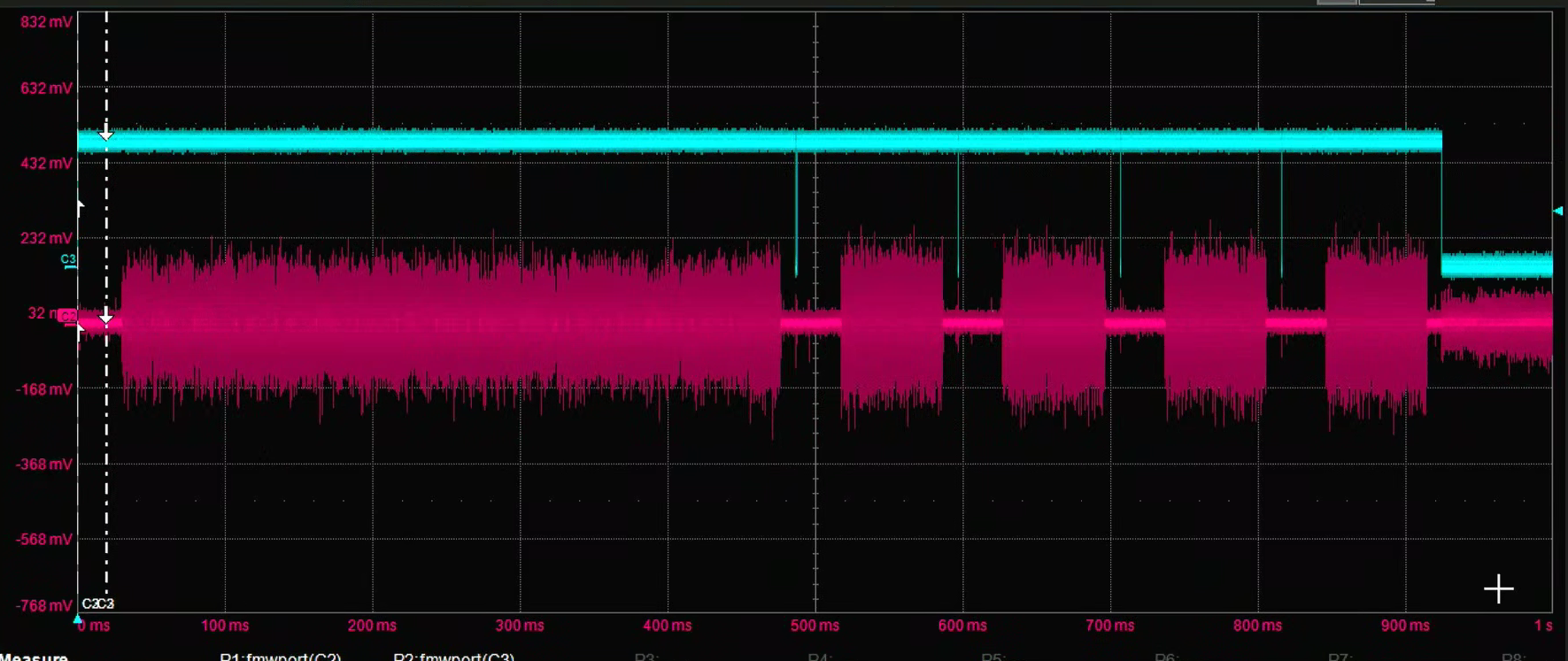}
  \caption{Oscilloscope capture showing one large prefill window followed by four repeating decode windows, corresponding to five output tokens in total.}
  \label{fig:setup-prefill-decode}
\end{figure}

An initial observation of the EM trace in Figure~\ref{fig:setup-prefill-decode} shows the prefill phase, followed by four decode phases.

\begin{figure}[b]
  \centering
  \includegraphics[width=\linewidth]{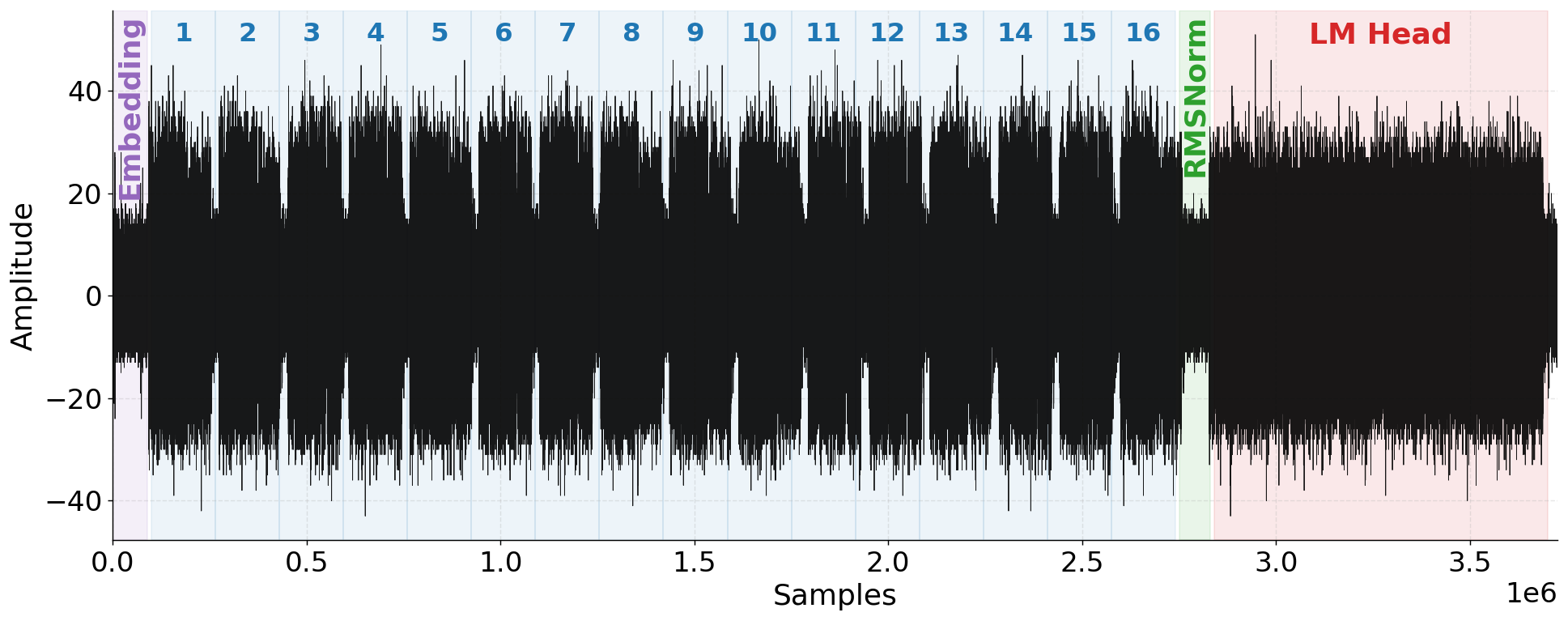}
  \caption{EM trace of a single generated token for LLaMA~3.2~1B Instruct. The execution flow is clearly visible in the physical emanations: it starts with the embedding layer, followed by the 16 transformer layers, the final RMSNorm, and concludes with the targeted LM Head projection.}
  \label{fig:execution-flow}
\end{figure}

Furthermore, we examine the EM trace of a single generated token at a high sampling resolution to map the physical emanations directly to the LLM's execution flow. Figure~\ref{fig:execution-flow} shows a distinct structural pattern in the EM trace that corresponds to the model's architecture. The computation begins with the embedding layer lookup, proceeds through the sequential execution of the 16 transformer blocks of LLaMA~3.2~1B, applies the final RMSNorm, and concludes with the LM Head matrix-vector multiplication. We leverage this clear temporal isolation to pinpoint the specific computation window corresponding to the LM Head for targeted side-channel analysis.

\subsection{Leakage Validation}
\label{sec:leakage-validation}

\begin{figure}[htbp]
  \centering
  \includegraphics[width=\linewidth]{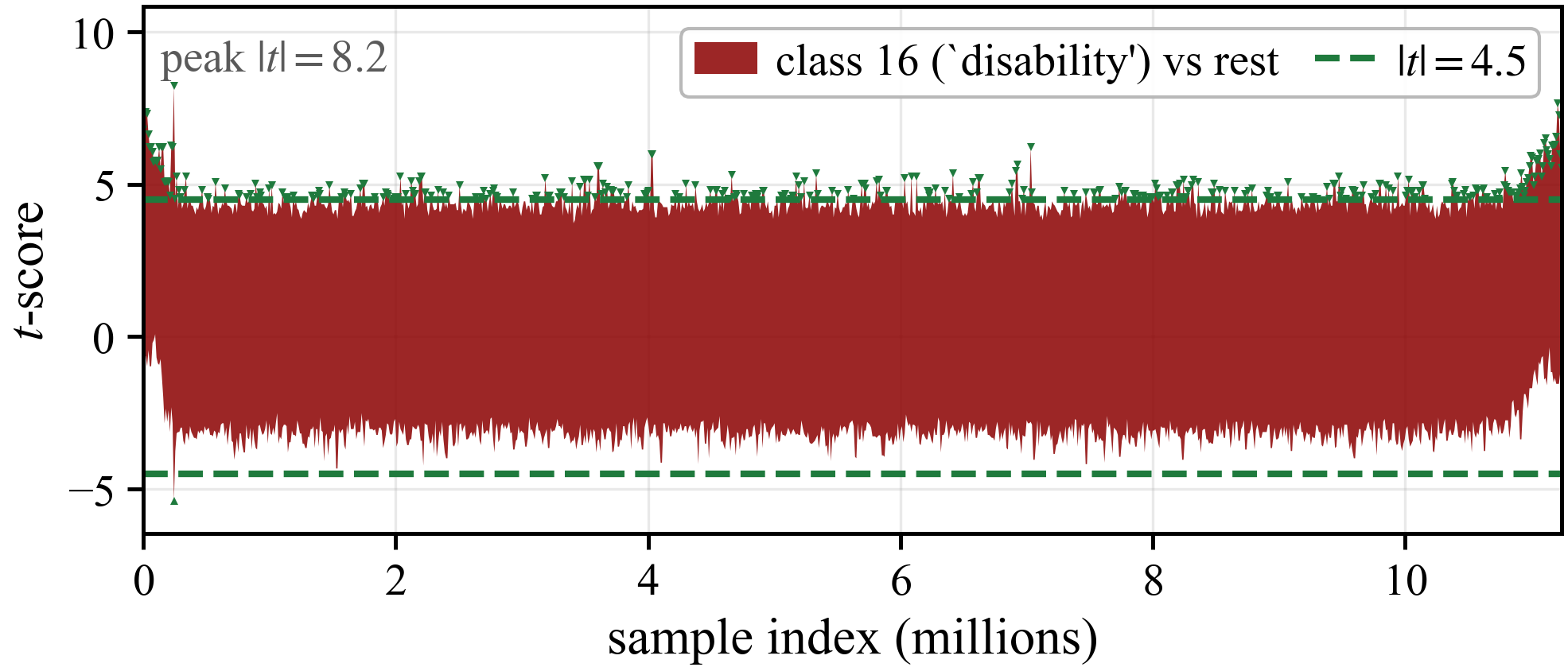}
  \caption{TVLA $t$-statistic for class 16 ('disability') against all other classes. The statistic clearly exceeds the $|t|=4.5$ threshold (green dashed line), demonstrating a statistically detectable difference between the trace populations during the LM Head computation.}
  \label{fig:tvla}
\end{figure}

Before building an attack model, we apply test vector leakage assessment (TVLA) to verify that the captured EM measurements exhibit statistically detectable differences between tokens. We estimate this using Welch's $t$-statistic, computed at every sample of the window. Following standard side-channel literature, we consider the measurement set to exhibit statistically detectable leakage when the absolute $t$-statistic exceeds a threshold of 4.5 at any sample. As Figure~\ref{fig:tvla} shows, our measurements consistently exceed this threshold, demonstrating a statistically detectable difference between the trace populations.

To identify the best way to extract leakage, this study evaluates three input representations: the raw 1D EM trace, a short-time Fourier transform (STFT) spectrogram, and a Morlet wavelet scalogram~\cite{grossmann1984decomposition}. The raw 1D trace contains all captured information, but processing millions of raw samples directly is computationally expensive and can hide frequency-localized leakage. The two time--frequency representations expose both the time and frequency structure to the convolutional networks, and allow the use of standard computer vision models pretrained on large datasets. 
We evaluate the STFT as our first candidate, which uses a fixed sliding window to map the signal into a uniform grid of time and frequency bins. 
Alternatively, we evaluate the Morlet wavelet scalogram. Unlike the STFT, the wavelet transform stretches into a long window to measure low frequencies and compresses into a short window to measure high frequencies \cite{sao2024joint, raihan2005wavelet}. This approach requires more computation, but it helps resolve short transient events that the STFT's fixed window may miss.

Based on the pilot study evaluated in Section~\ref{ch:results}, the final pipeline utilizes STFT spectrograms resized to $256 \times 256$ pixels. To accommodate pretrained vision backbones, the single-channel spectrogram is replicated across the three input channels, without applying a color map, as this configuration provided the highest feature extraction accuracy.


\begin{figure}
  \centering
  \includegraphics[width=1\linewidth]{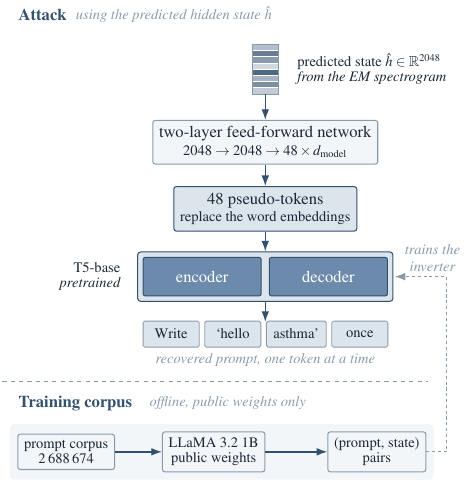}
    \caption[Semantic prompt recovery]{Semantic recovery of the submitted prompt from the predicted hidden state. The upper path depicts the attack: the state is passed through a two-layer MLP into 48 pseudo-tokens that replace word embeddings at the encoder input. The decoder generates the prompt one token at a time. The bottom path represents the training corpus, which is produced offline by running prompts through the public model and recording the resulting states.}
  \label{fig:prompt-reconstruction}
\end{figure}

\subsection{Semantic Prompt Recovery}
\label{sec:method-prompt-inversion}
The hidden state that generates a token also encodes the entire input context. In language models, the self-attention mechanism ensures that the representation at any given step aggregates information from all preceding tokens. As a result, this single continuous vector acts as a summary of the entire prompt. To reconstruct the user prompt, we apply a semantic prompt recovery attack by adapting embedding inversion techniques~\cite{morris2023vec2text,morris2024languagemodelinversion} to operate directly on the continuous hidden state. As shown in Figure~\ref{fig:prompt-reconstruction}, the recovery model uses a sequence-to-sequence architecture. We use a two-layer feed-forward network to project the single 2\,048-dimensional hidden state vector into $48 \times d_{\text{model}}$ values, which we reshape into a sequence of 48 vectors. These serve as input embeddings for a pretrained T5-base encoder-decoder~\cite{raffel2020t5}, which we train via cross-entropy to generate the original prompt token by token. 

Because the LLaMA~3.2~1B weights are public, we generate the required training dataset of 2.68 million prompt-state pairs in software, introducing zero physical profiling overhead. We utilize public datasets of instructions, question-answering, conversations, and domain-specific texts (detailed in Table~\ref{tab:corpus-sources}) to source the majority of the prompts. We then process each prompt through the model to capture the resulting hidden state. Of the 2\,688\,674 pairs, we use roughly 
95\% for training, and hold out the remainder for evaluation. Utilizing a diverse set of sources ensures breadth of phrasing; the attack must recover prompts written in language the model has never seen. We completely exclude all 17 prompt templates used during EM trace collection from the training corpus to force the model to semantically recover unseen phrasing.

\section{Results}
\label{ch:results}

\subsection{Closed-Set Token Recovery}
\label{sec:leakage-characterization}

This section evaluates the performance of the proposed attacks. Unless stated otherwise, the results are derived from models trained and evaluated on 50 classes, 120\,750-trace corpus described in Section~\ref{ch:model} (84\,524 training, 18\,113 validation, and 18\,113 test traces), using single capture Template~\ref{quote:template}, and following the preprocessing pipeline from Section~\ref{ch:model}. We first identify an optimal input representation and network architecture using a smaller 30-class subset. This pilot study reduces the computational cost of the architecture search, and we train the best-performing configuration on the full corpus. Subsequently, we present the results of the 50-class attack, assessing token classification accuracy, the location of leakage within the trace, data requirements, and the extension from a fixed vocabulary to token recovery over the full vocabulary.

\label{sec:pilot}
\noindent\textbf{Input Representation and Network Architecture.}
We evaluate one-dimensional classifiers (MLPs, 1D CNNs, TCNs, 1D ResNets) trained directly on the raw trace. However, as shown in Table~\ref{tab:final}, none rise meaningfully above chance at any capacity, so the primary classification models are built on two-dimensional representations. 

For the time-frequency images, we evaluate two families of models. The first is a custom residual network trained from scratch, which isolates the contribution of pretrained weights. It projects to 50 classes. The second family fine-tunes image backbones pretrained on ImageNet~\cite{deng2009imagenet}: ResNet-18, ResNet-50~\cite{he2016resnet}, and EfficientNet-B0~\cite{tan2020efficientnetrethinkingmodelscaling}. This tests whether features learned from natural images transfer effectively to EM spectrograms. The final classification layer of each backbone is replaced to match the target output.

We train the classification models with cross-entropy loss, label smoothing $0.1$~\cite{szegedy2016rethinking}, Adam optimizer~\cite{kingma2015adam} with weight decay $10^{-3}$, and a cosine learning-rate decay schedule. Early stopping selects the checkpoint with the lowest validation loss after up to 40 epochs.

Table~\ref{tab:pilot} presents the results of the 30-class pilot study, which evaluates 20 distinct configurations. These experiments compared five input representations paired with various compatible network architectures. The pilot uses the same capture, preprocessing, and training pipeline as the full task, with a smaller subset to keep the search affordable. The classes are balanced, with a random guessing accuracy of 3.3\%.

\begin{table}[t]
  \centering
  \caption[Representation and architecture pilot, 30 classes]{The 30-class pilot, sorted by top-1 accuracy. \emph{repl.} replicates the single-channel map across the three input channels; \emph{cmap} applies a color map instead.}
  \label{tab:pilot}
  \small
  \resizebox{\linewidth}{!}{
  \begin{tabular}{@{}llllrr@{}}
    \toprule
    Representation & Size & Channels & Architecture & Top-1 \% & Top-5 \% \\
    \midrule
    STFT      & $256^2$ & repl. & pretrained ResNet-18   & 98.28 & 99.97 \\
    STFT      & $224^2$ & repl. & pretrained ResNet-18   & 98.14 & 99.99 \\
    STFT      & $224^2$ & cmap  & pretrained ResNet-18   & 98.01 & 99.97 \\
    STFT      & $224^2$ & repl. & pretrained ResNet-50   & 97.73 & 99.96 \\
    STFT      & $224^2$ & repl. & fine-tuned CLIP RN-50  & 97.37 & 99.83 \\
    STFT      & $224^2$ & repl. & pretrained EffNet-B0   & 92.35 & 99.81 \\
    STFT      & $256^2$ & repl. & pretrained EffNet-B0   & 91.69 & 99.83 \\
    STFT      & $224^2$ & cmap  & pretrained EffNet-B0   & 91.61 & 99.81 \\
    STFT      & $224^2$ & cmap  & pretrained ResNet-50   & 75.77 & 97.99 \\
    STFT      & $256^2$ & repl. & pretrained ResNet-50   & 67.58 & 92.38 \\
    Scalogram & $224^2$ & repl. & pretrained ResNet-18   & 53.70 & 91.20 \\
    Scalogram & $224^2$ & repl. & pretrained ResNet-50   & 50.94 & 89.68 \\
    Scalogram & $224^2$ & repl. & pretrained EffNet-B0   & 35.45 & 80.29 \\
    Scalogram & native  & 1ch   & from scratch           & 23.86 & 70.76 \\
    STFT      & $256^2$ & repl. & from scratch           & 22.18 & 75.98 \\
    STFT      & $224^2$ & cmap  & from scratch           & 20.00 & 69.62 \\
    Scalogram & $224^2$ & repl. & from scratch           & 15.31 & 62.50 \\
    STFT      & $224^2$ & repl. & fine-tuned CLIP ViT-B/32    & 7.34  & 36.31 \\
    STFT      & $224^2$ & repl. & fine-tuned MAE ViT-B/16         & 7.31  & 36.52 \\
    STFT      & $224^2$ & repl. & from scratch           & 7.21  & 35.87 \\
    \bottomrule
  \end{tabular}
  }
\end{table}

The three pretrained backbones (ResNet-18~\cite{he2016resnet}, ResNet-50~\cite{he2016resnet}, and EfficientNet-B0~\cite{tan2020efficientnetrethinkingmodelscaling}) are ImageNet~\cite{deng2009imagenet} models of different families. The model trained from scratch is the control without ImageNet weights, so the difference between them measures the effect of pretraining.

Three main observations emerge from the table. First, the STFT spectrogram outperforms the scalogram across all resolutions and architectures: the pretrained ResNet-18 achieves 98.1\% accuracy on the $224\times224$ STFT compared to 53.7\% on the corresponding scalogram. Second, pretraining improves performance: the best from-scratch configuration, 23.9\%, is less accurate than the weakest pretrained model, 50.9\%, indicating that the training data in the 30-class pilot is insufficient to train a network from random initialization.

ResNet-18 is the most consistent architecture. ResNet-50 remains competitive on the replicated spectrogram, but its accuracy decreases on the color-mapped input 75.8\% and the $256\times256$ input 67.6\%. ResNet-18 and EfficientNet-B0 do not exhibit this sensitivity.

The transformer-based architectures struggle to adapt to the side-channel data. Despite fine-tuning, both the CLIP ViT-B/32 and the Masked Autoencoder (MAE) ViT-B/16 remain near 7\% accuracy. In contrast, the CLIP ResNet-50 backbone fine-tunes to 97.37\% accuracy, indicating that the convolutional inductive bias is highly beneficial for processing EM spectrograms. Ultimately, the $256\times256$ STFT spectrogram paired with the ImageNet-pretrained ResNet-18 yields the highest overall accuracy. This configuration is carried forward to the 50-class corpus for the rest of the section.

\label{sec:full50}
\noindent\textbf{50-Class Classification Performance.}
On the 50-class corpus, the $256\times256$ STFT with the pretrained ResNet-18 achieves 97.2\% top-1 and 99.95\% top-5 accuracy on the test set, as shown in Table~\ref{tab:final}, against 98.3\% for the same configuration on the 30-class pilot. The training process converges without signs of overfitting; the validation loss decreases steadily until the end of training, which provides the best checkpoint (epoch 39). Furthermore, the gap between training and validation accuracy is less than two percentage points.



\begin{table}[htbp]
  \centering
  \caption[Spectrogram model against raw 1D representations]{The highest-accuracy image model (the $256^2$ STFT spectrogram with a pretrained ResNet-18) against raw 1D models of different capacities, all trained and evaluated on the same 50-class corpus.
  The raw-trace models isolate the effect of the input format. Chance is 2\%.}
  \label{tab:final}
  \small
  \begin{tabular}{@{}llrr@{}}
    \toprule
    Model & Params & Top-1 \% & Top-5 \% \\
    \midrule
    STFT $256^2$ + pretrained ResNet-18 & 11.2M & 97.20 & 99.95 \\
    \midrule
    Raw 1D CNN            & 0.19M & 6.56 & 31.70 \\
    Raw 1D CNN (wide)     & 1.30M & 6.13 & 30.45 \\
    Raw 1D CNN (deep)     & 6.69M & 6.07 & 30.02 \\
    Raw 1D ResNet         & 11.3M & 6.03 & 29.82 \\
    Raw 1D TCN            & 0.61M & 6.00 & 30.04 \\
    Raw 1D TCN (deep)     & 3.73M & 6.00 & 29.97 \\
    Raw 1D MLP            & 577M  & 5.58 & 28.62 \\
    Raw 1D MLP (large)    & 1.16B & 5.26 & 28.32 \\
    \bottomrule
  \end{tabular}
\end{table}


Table~\ref{tab:final} compares the spectrogram model with raw 1D models that receive the same signal without the time-frequency transform. The raw 1D models perform only slightly better than random guessing, maintaining an accuracy of 5--7\% across all tested sizes. Increasing model capacity, whether by widening or deepening the convolutional network, employing a temporal convolutional network, or scaling a multi-layer perceptron to 1.16 billion parameters, does not improve accuracy. Conversely, the same signal achieves 97.2\% accuracy when converted to a spectrogram. The sharp increase in accuracy implies that the necessary information is present in the trace, but a raw 1D network is unable to extract it over this sequence length. In contrast, the time-frequency representation successfully exposes it. 

\begin{figure}
  \centering
  \includegraphics[width=0.8\linewidth]{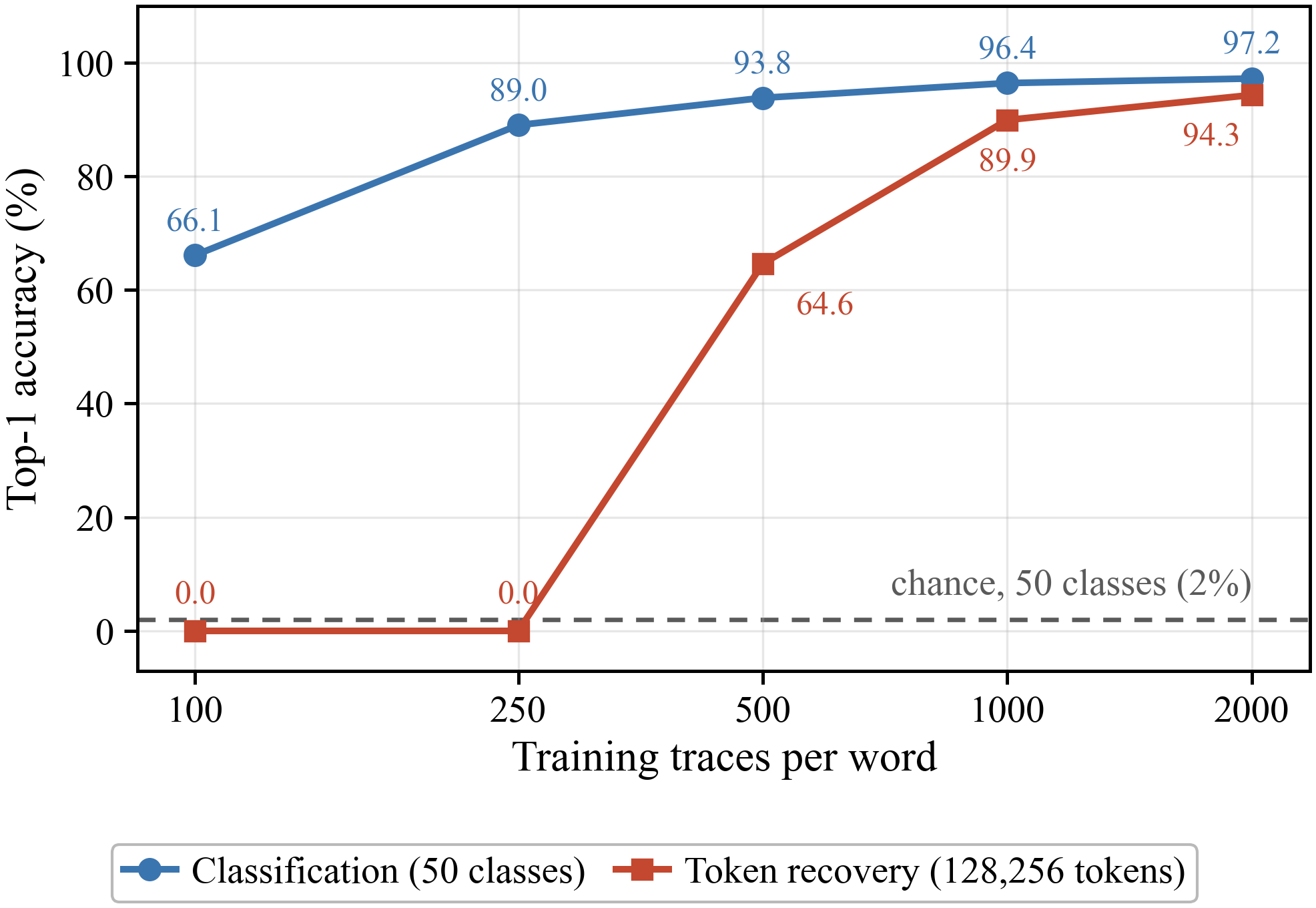}
  \caption[Training traces per word]{Top-1 accuracy against training traces per word. Classification is usable from a few hundred traces per word, whereas the generative model stays at chance until 500 and needs roughly four times as much data to reach the same accuracy.}
  \label{fig:trace-budget}
\end{figure}

\label{sec:trace-budget}
\noindent\textbf{Training Data Requirements.}
The number of traces an attacker must collect per word determines how much physical access the attack requires, and is therefore the main practical cost of the profiling stage. This section measures the cost for both attacks by training each on a fixed number of traces per word and evaluating on the same test set. The sweep covers 100 to 2\,000 traces per word. Table~\ref{tab:budget} reports accuracy at each budget, and Figure~\ref{fig:trace-budget} shows the same data as learning curves. Both attacks use the 50-word corpus: classification over the 50 known words, and the generative model is scored by top-1 token recovery over the full 128\,256-token vocabulary.

\begin{table}[htbp]
  \centering
  \caption[Accuracy against training traces per word]{Top-1 accuracy at each training budget, on the 50-word corpus. Classification is closed-set over fifty words; the generative model is scored over all 128\,256 tokens. Chance is 2\% and $7.8\times10^{-6}$ respectively.}
  \label{tab:budget}
  \small
  \begin{tabular}{@{}rrr@{}}
    \toprule
    Traces/word & Classification (\%) & Generative (\%) \\
    \midrule
    100  & 66.1 & 0.0 \\
    250  & 89.0 & 0.0 \\
    500  & 93.8 & 64.6 \\
    1000 & 96.4 & 89.9 \\
    2000 & 97.2 & 94.3 \\
    \bottomrule
  \end{tabular}
\end{table}

The two attacks have different data requirements. Classification reaches 89.0\% with 250 traces per word and 93.8\% with 500. With 1000 traces, the accuracy increases by 2.6\%; going from 1000 traces to 2000 increases accuracy by 0.8\%. Collecting a few hundred traces per word is therefore enough for the closed-set attack. The generative model requires more data. It remains at chance below 500 traces per word, reaches 64.6\% at 500, and 94.3\% at 2\,000. The two tasks differ in what they predict: the classifier chooses among 50 labels, while the generative model predicts a 2\,048-dimensional continuous state that is then decoded offline against the full vocabulary. The second is the harder estimation problem, and the curves reflect that. 

\label{sec:leakage}
\noindent\textbf{Localizing the Leakage.}
To identify the regions of the EM trace that the model uses for classification, and thereby to locate the potential source of the leakage, parts of the trace or spectrogram are masked, and the subsequent decrease in accuracy is recorded.

\begin{figure}[t]
  \centering
  \includegraphics[width=0.7\linewidth]{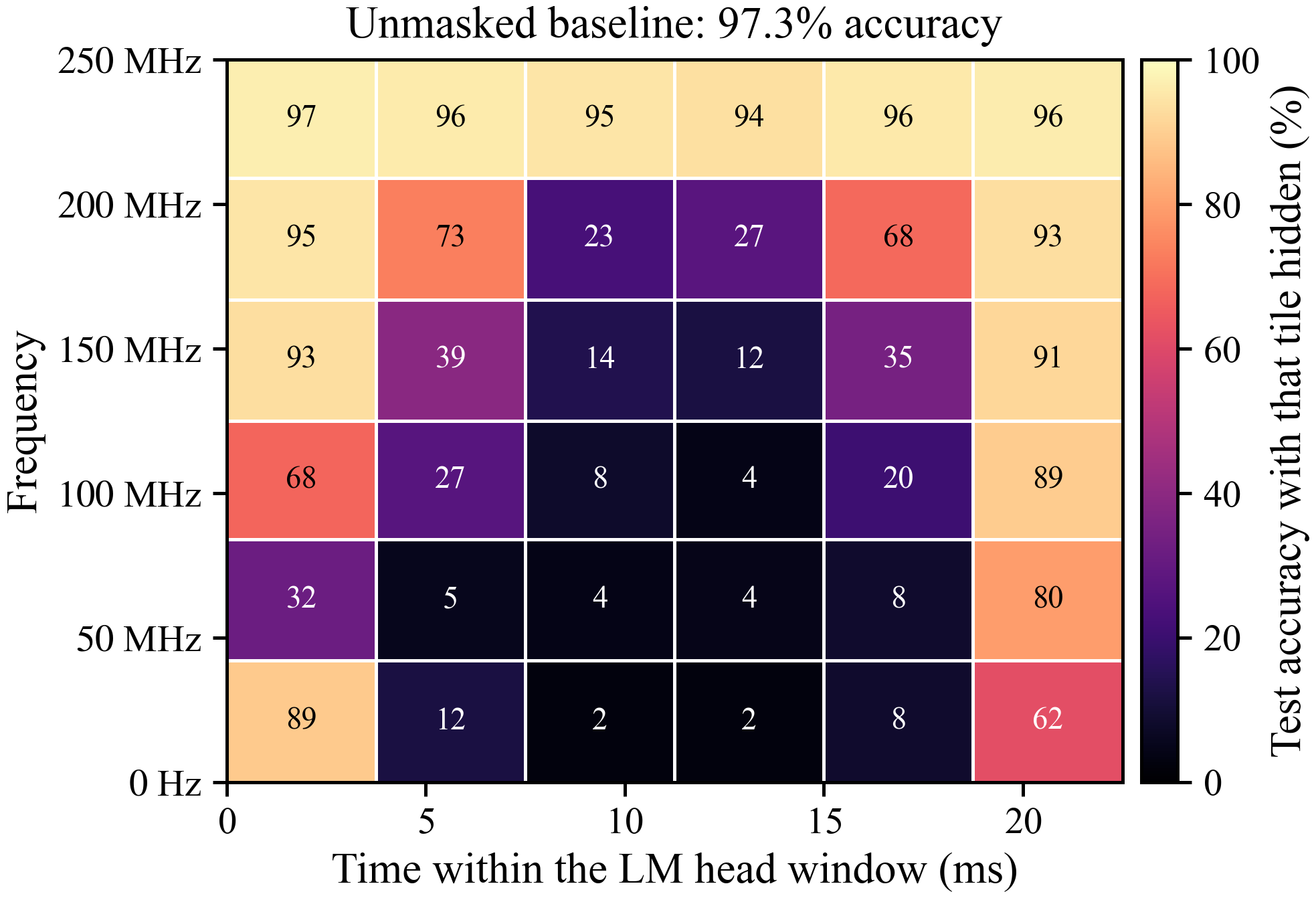}
  \caption[$6\times6$ block mask]{Accuracy when one time-frequency tile of a $6\times6$ grid is hidden at a time.}
  \label{fig:block-mask}
\end{figure}

To isolate the leakage in both time and frequency simultaneously, a $6\times6$ grid mask (Figure~\ref{fig:block-mask}) separates the time and frequency axes by hiding one time-frequency tile at a time. The top frequency row 200--250~MHz provides little information across the LM Head window, and masking it has minimal effect on accuracy. The leakage is concentrated in the middle third of the LM Head window in time and 0--150~MHz in frequency, where masking a single tile reduces accuracy to near zero. In contrast, masking tiles at the beginning or end of the LM Head window has minimal impact, even at low frequencies.

\begin{figure}
  \centering
  \includegraphics[width=0.8\linewidth]{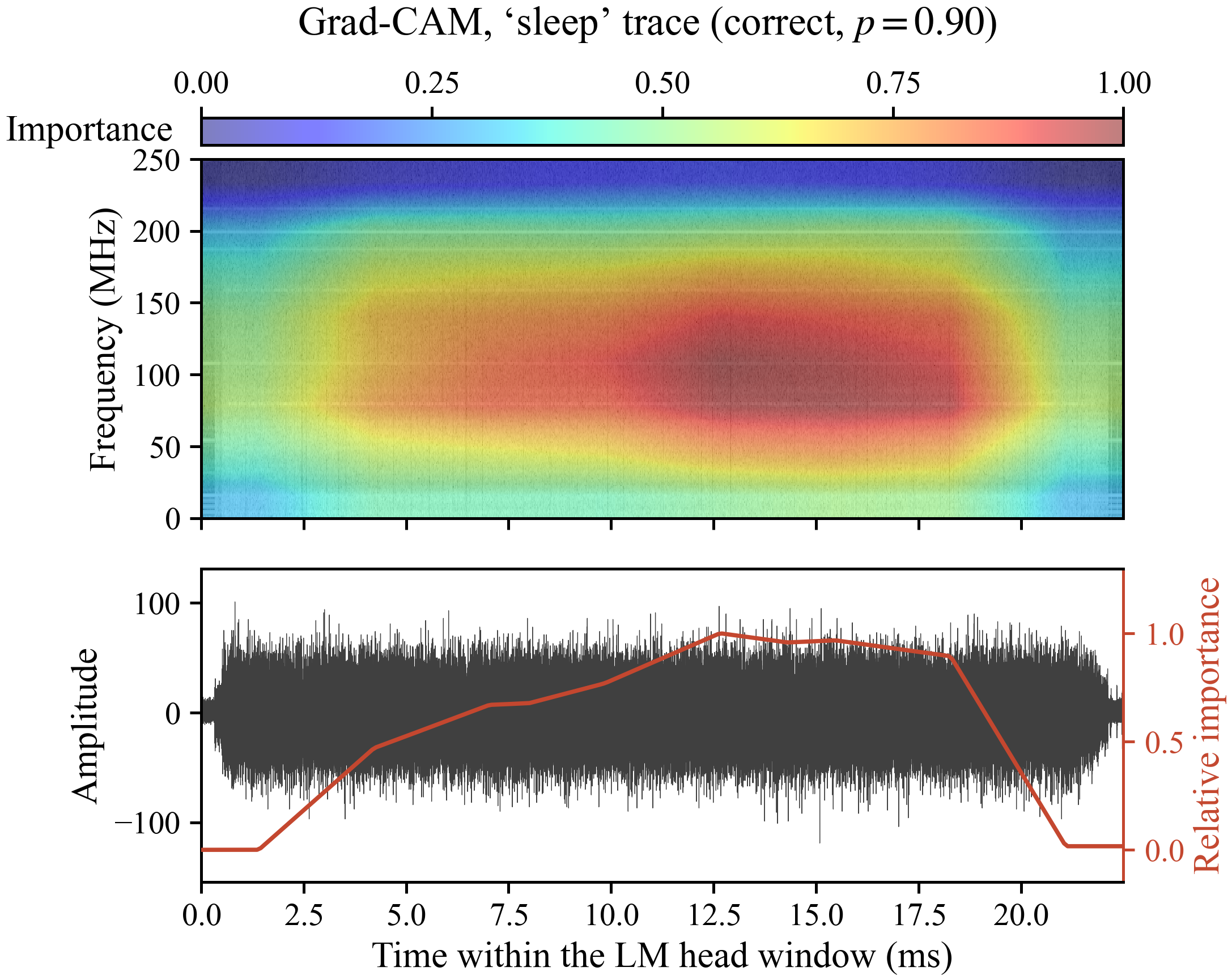}
  \caption{Grad-CAM attribution for a test trace, shown on the resized input, native spectrogram, and back-projected onto the raw time-domain trace, highlighting the same region identified by the grid mask.
  }
  \label{fig:gradcam}
\end{figure}

To further analyze the leakage, we apply Grad-CAM~\cite{Selvaraju_2019}, a gradient-based method that highlights the input regions that are most influential in the model's prediction. We project the resulting heatmaps onto the native spectrogram and raw time-domain trace in Figure~\ref{fig:gradcam}. As expected, Grad-CAM highlights the same mid-time, low-to-mid-frequency region identified by the block mask in Figure~\ref{fig:block-mask}. Despite their distinct mechanisms, both methods isolate the same leakage region.

\label{sec:domain}
\noindent\textbf{Limitations of Closed-Set Classification.}
While the classifier accurately identifies tokens from the profiled 50-word vocabulary, a deployed LLM generates arbitrary text. A fixed-capacity classifier lacks a mechanism to reject out-of-vocabulary (OOV) inputs and is forced to map every trace to one of its known classes. 

\begin{figure}
  \centering
  \includegraphics[width=\linewidth]{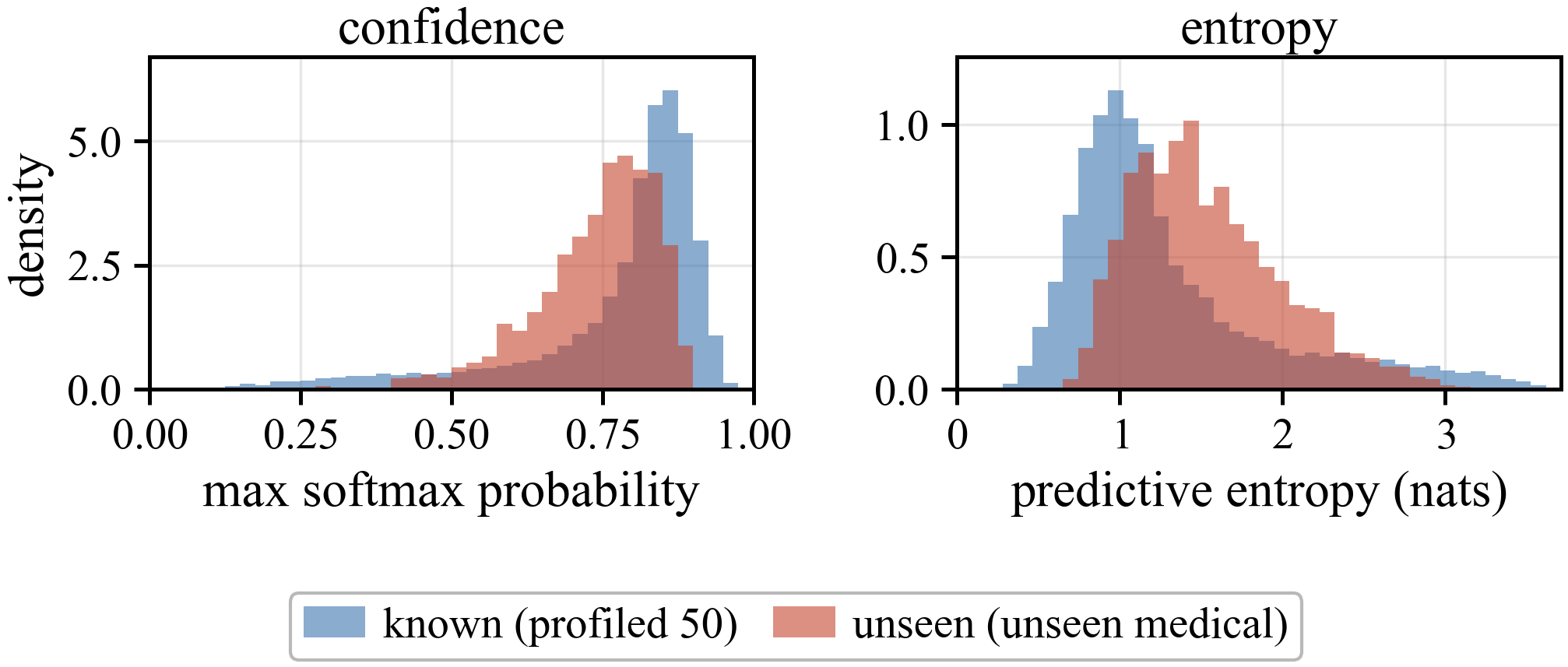}
  \includegraphics[width=\linewidth]{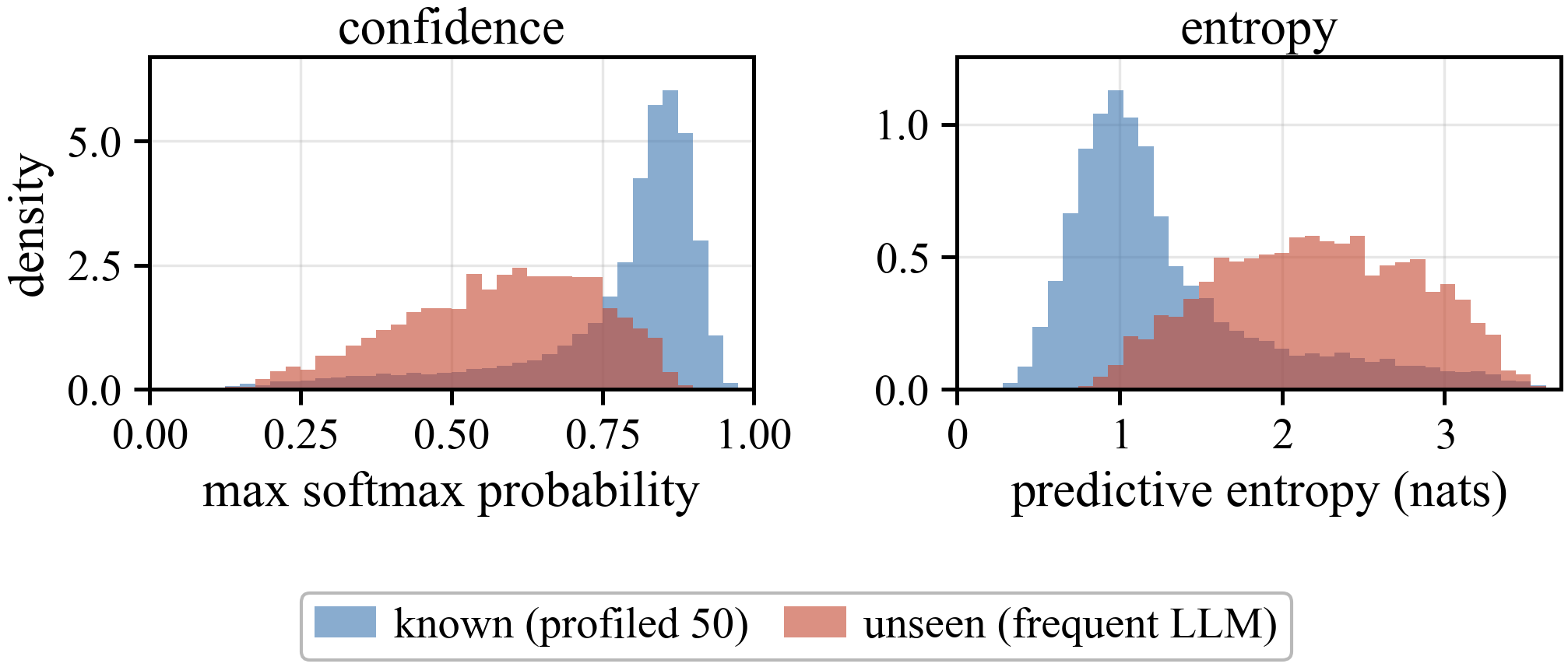}
  \includegraphics[width=\linewidth]{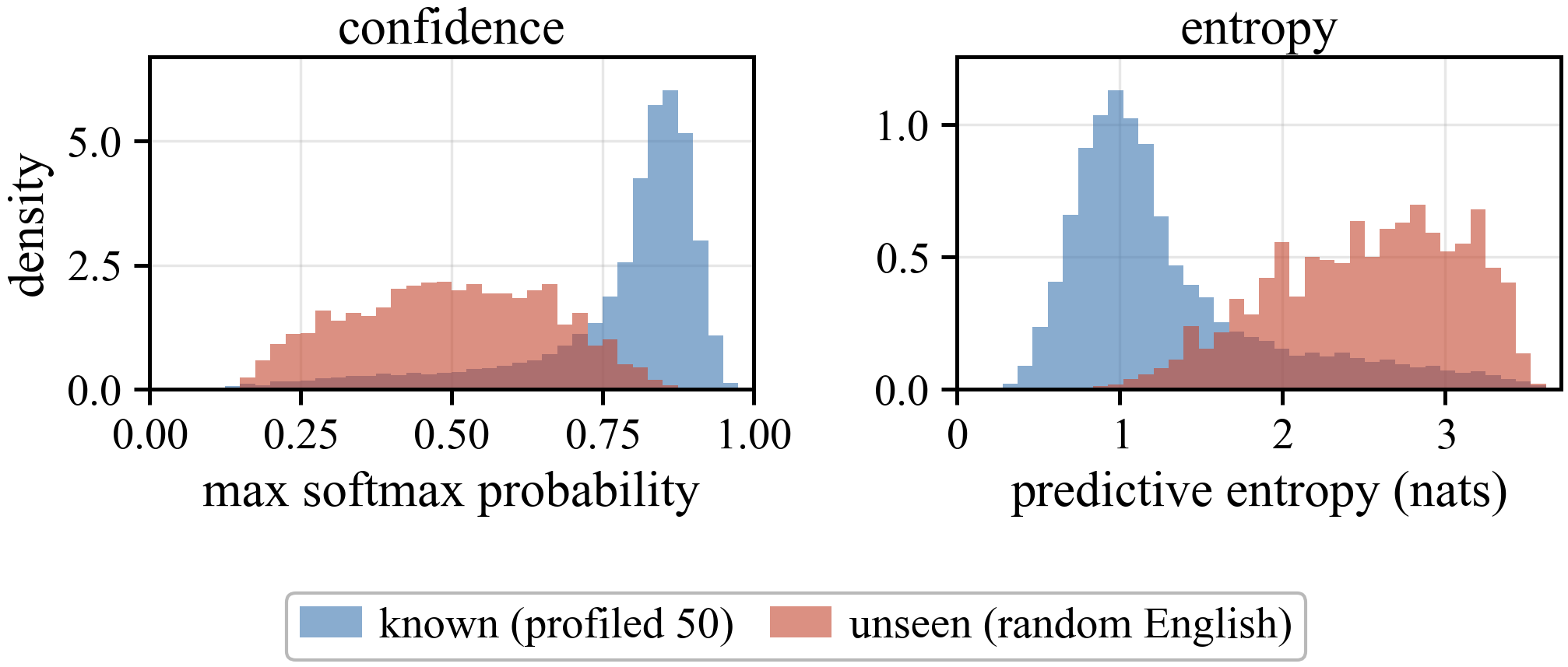}
  \caption[Out-of-vocabulary confidence distributions]{Maximum softmax probability the 50-class classifier assigns to unseen tokens from three groups: unseen medical words (top), frequent LLM words (middle), and random English words (bottom). The classifier stays confident on out-of-vocabulary words, with most confidence on the in-domain medical words.}
  \label{fig:openset-conf}
\end{figure}

To evaluate OOV behavior, the 50-class classifier was tested on traces of unprofiled words sampled from three categories: unseen medical terms, frequent LLM words, and random English words. As shown in Figure~\ref{fig:openset-conf}, random and frequent words exhibit lower confidence, but unseen medical words maintain high confidence distributions that heavily overlap with known (profiled) words due to domain similarity. Consequently, even without identifying the specific word, an attacker observing high confidence on an unrecognized token can infer that the user is discussing a medical topic, which constitutes a privacy disclosure in itself. 

Because closed-set classification confidently maps unfamiliar words to existing classes and is constrained from scaling to the full output of an LLM, recovering arbitrary text necessitates a generative approach.

\label{sec:recovery}
\subsection{Open-Vocabulary Token Recovery}
Closed-set classification is bounded by its predefined vocabulary. To address this limitation, the second task takes a generative approach to token recovery: a pretrained EfficientNet-B0 processes the $256\times256$ STFT spectrogram to predict the post-RMSNorm hidden state $\hat{h}\in\mathbb{R}^{2048}$. Given that the LLaMA~3.2~1B weights are public, we project this predicted hidden state to the vocabulary space using the LM Head projection formula defined in Equation~\ref{eq:lmhead-bg}. This produces logits $z = W\hat{h}$ over all 128\,256 tokens, allowing us to identify the most probable token without relying on a pre-profiled dictionary.

Recovery is evaluated directly on this decoded ranking. For each test trace, the generated logits are used to rank all 128\,256 tokens. Top-1 accuracy denotes the fraction of traces where the true token ranks first, while the median rank summarizes overall ranking performance. 

\vspace{0.5ex}\noindent\textbf{Architecture and Objective Selection.}
To determine the optimal configuration for hidden-state prediction, we conducted a pilot architecture search using only the profiled words on a reduced budget of 1\,000 traces per class. Pretrained CNNs (EfficientNet-B0, ResNet-18, and ResNet-50) effectively solved the task, achieving a median rank of one, with EfficientNet-B0 demonstrating the strongest performance. Conversely, models trained from scratch and transformer-based architectures (CLIP ViT and MAE) performed near chance. For the regression objective, maximizing the cosine similarity between the predicted and target states yielded the highest accuracy (89.9\% top-1) compared to InfoNCE or Mean Squared Error. Consequently, the final generative model employs an EfficientNet-B0 backbone optimized exclusively with a cosine loss.

\vspace{0.5ex}\noindent\textbf{Recovery on Profiled Words.}
Trained on the full corpus with the cosine objective, the EfficientNet-B0 model reaches 94.3\% top-1 and 95.8\% top-5 accuracy over the whole 128\,256-token vocabulary, with a median rank of 1, so the predicted continuous state decodes to the correct token without a pre-profiled dictionary. 
Table~\ref{tab:budget} reports the model's performance at different trace budgets: the same configuration reaches 89.9\% accuracy with 1\,000 traces per word, and 94.3\% with 2\,000 traces per word.


This 94.3\% applies to \emph{profiled} words present in the training set. On unseen words, the same model places the correct token in the Top-100 for only 11.8\% of traces, with a median rank of 1\,327 out of 128\,256. The rank varies by domain, being lower for unseen words sharing the training domain (e.g., medical terms) and higher for out-of-domain words. The next two sections improve this generalization in two additive steps: centering the prediction target, then matching the training objective to the full-vocabulary decode. Together, they reduce the median rank from 1\,327 to 54 and raise Top-100 recovery from 11.8\% to 67.0\%.

\begin{figure}
  \centering
  \includegraphics[width=0.9\linewidth]{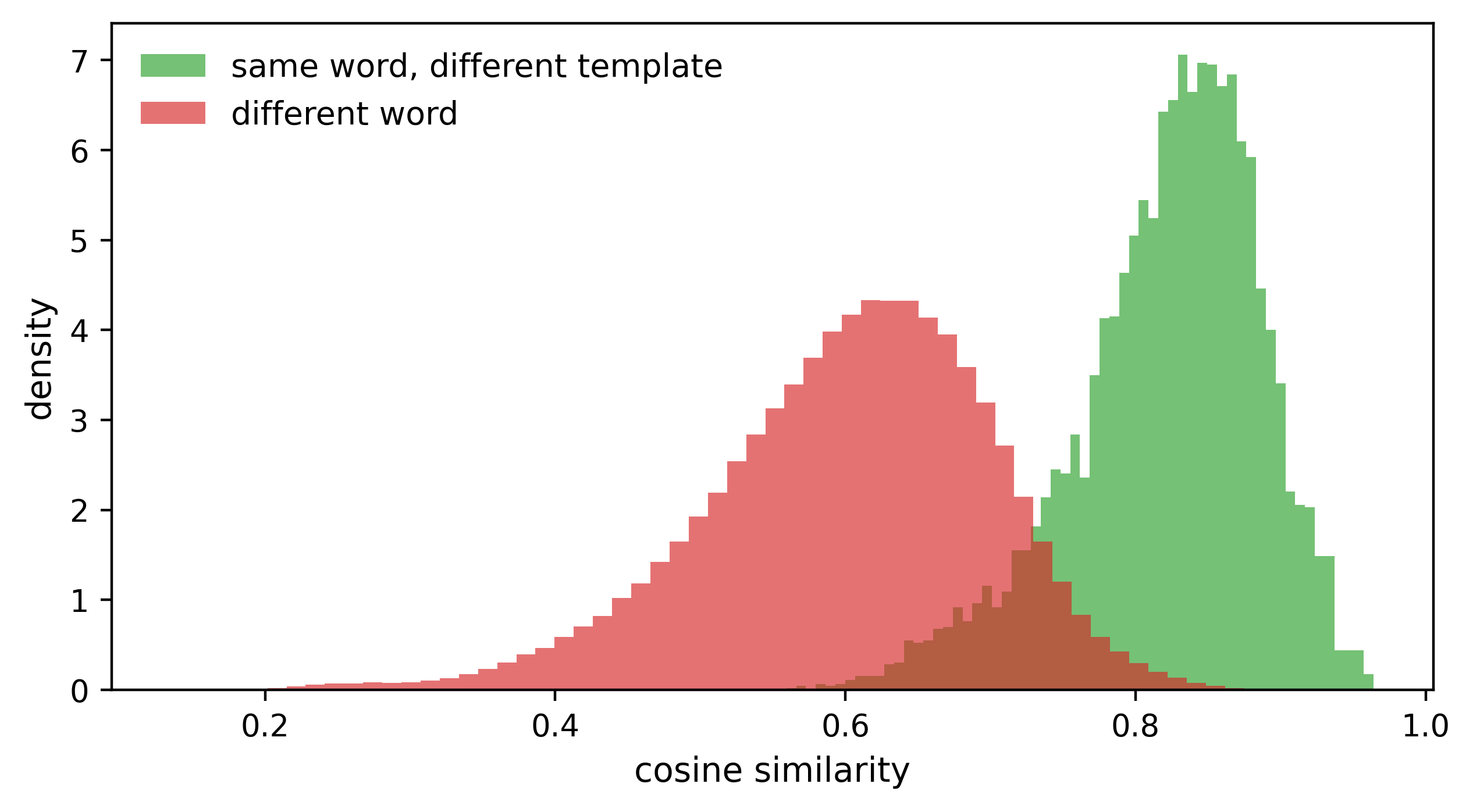}
  \caption[Hidden-state cosine similarity]{Density of pairwise cosine similarities over the 849 hidden states (50 words $\times$ 17 templates). The same word across different templates gives high similarity (mean 0.82); different words give lower similarity (mean 0.60, pooled over templates; Table~\ref{tab:word-template} splits this by template).}
  \label{fig:hidden-cosine}
\end{figure}

\label{sec:hidden-consistency}
\noindent\textbf{Robustness of the Hidden State to Prompt Variations.}
For full-vocabulary token recovery to succeed, the LLM's hidden state must be driven by the specific word being processed, rather than just the preceding prompt context. We evaluate this by analyzing the pairwise cosine similarities of 849 hidden states (50 words $\times$ 17 templates, minus one refused generation), as visualized in Figure~\ref{fig:hidden-cosine} and detailed in Table~\ref{tab:word-template}. 
The results show that the word is the dominant factor shaping the hidden state. Matching the word while the template differs yields a high similarity of 0.822, whereas matching the template while the word differs yields a lower similarity of 0.712 (the baseline is 0.596 when both differ). This indicates the word contributes slightly more to the representation space than the template. 
\begin{table}[htbp]
  \centering
  \caption[Word versus template in the hidden state]{Mean pairwise cosine between the 849 hidden states, grouped by whether the word and the template match. The same word under the same template is a single deterministic vector, so that cell has no off-diagonal pairs.}
  \label{tab:word-template}
  \small
  \begin{tabular}{@{}llr@{}}
    \toprule
    Word & Template & Mean cosine \\
    \midrule
    same      & different & 0.822 \\
    different & same      & 0.712 \\
    different & different & 0.596 \\
    \bottomrule
  \end{tabular}
\end{table}

\begin{figure}
  \centering
  \includegraphics[width=\linewidth]{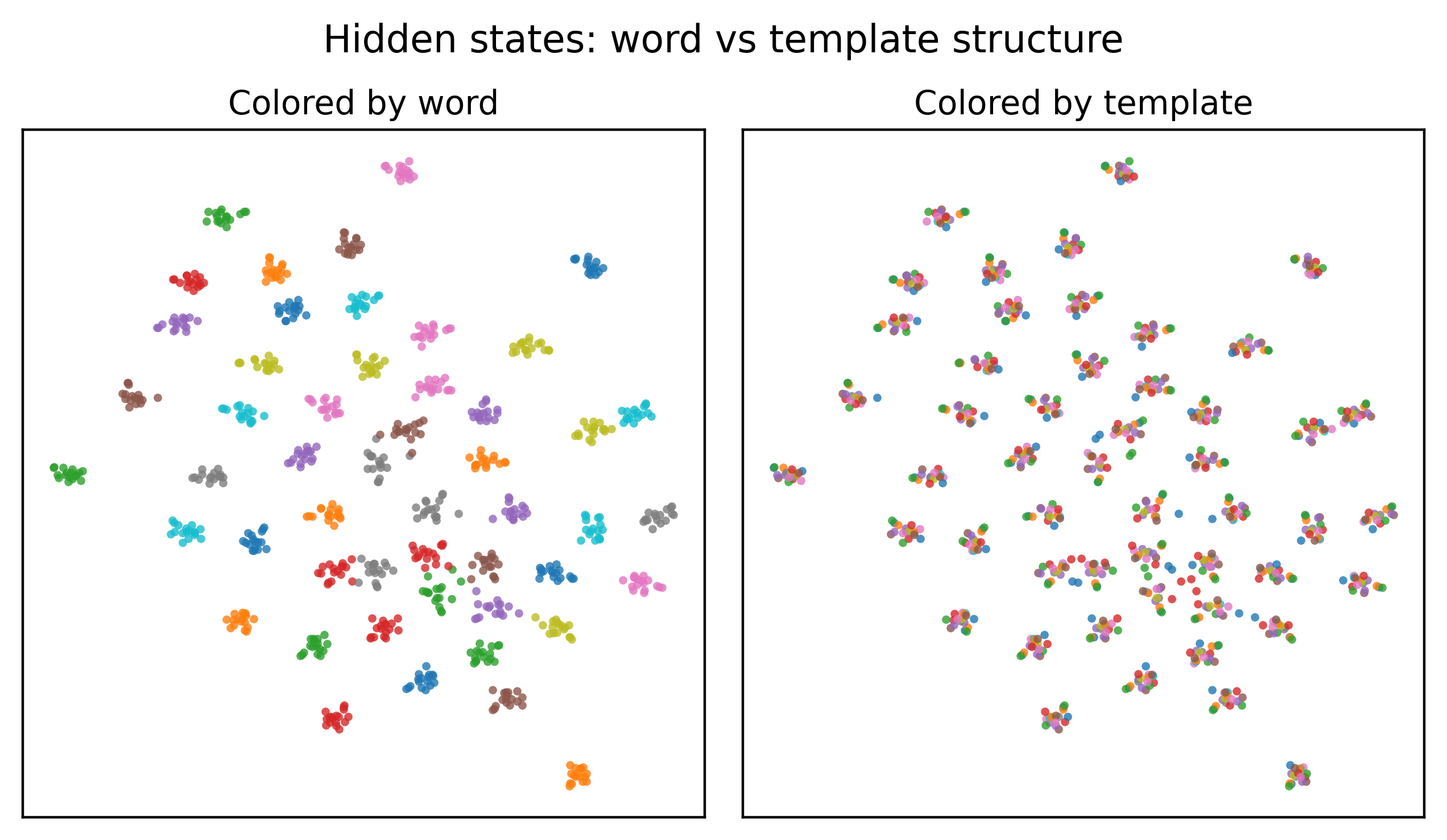}
  \caption[t-SNE of hidden states]{t-SNE projection of the LLM hidden states, colored by word (left) and template (right). Distinct clusters emerge when colored by word but not by template, indicating that the representation is primarily determined by the word rather than the surrounding context.}
  \label{fig:hidden-tsne}
\end{figure}

A t-SNE projection in Figure~\ref{fig:hidden-tsne} confirms this structure: hidden states form clear, tight clusters by word, and no clustering occurs by template. 

\label{sec:target-centering}
\vspace{0.5ex}\noindent\textbf{Isolating the Shared Component for Unseen-Word Recovery.}
Although the baseline model successfully decodes profiled words across the full vocabulary, generalizing to unseen tokens presents an additional challenge. The predicted hidden states for unprofiled inputs initially show less alignment with their true targets. We find that the target hidden states share a large common component; distinct words exhibit a mean cosine similarity of 0.596, as Table~\ref{tab:word-template} shows. A recovery model trained on the raw state $h$ primarily learns this shared direction, which can overshadow the word-specific components required to resolve unseen inputs. To isolate these discriminative features, we center the target by predicting the residual $r = h - \mu$ and restore the mean $\mu$ prior to offline decoding ($\hat{h} = \hat{r} + \mu$). As Table~\ref{tab:centering-similarity} demonstrates, this transformation reduces the mean similarity between distinct words to near zero and widens the separation between traces of the same word and different words. The residual target remains fully decodable, ranking the true token first for 92.9\% of profiled states. This approach preserves the necessary information while directing the network to model the word-specific geometry.

\begin{table}[htbp]
  \centering
  \caption[Target centring cosine similarities]{Mean cosine similarities between traces of the same word (within-word) and different words (between-word) before and after target centring.}
  \label{tab:centering-similarity}
  \begin{tabular}{@{}lccc@{}}
    \toprule
    Target & Within-word & Between-word & Gap \\
    \midrule
    Raw $h$ & $+0.822$ & $+0.602$ & 0.219 \\
    Centred $r$ & $+0.551$ & $-0.011$ & 0.562 \\
    \bottomrule
  \end{tabular}
\end{table}

As shown in Table~\ref{tab:centering}, target centering alone improves recovery for unseen words. Specifically, the median rank improves from 1\,327 to 443, with the fraction of traces ranking the true token within the top 100 rising from 11.8\% to 28.7\%. Top-1 accuracy for unseen words remains at 0\%; however, centering shifts the prediction from generic high-frequency tokens (e.g., \texttt{I}, \texttt{the}) toward the semantic neighborhood of the true word. For instance, predictions for the unseen word \emph{clinic} shift to highly related terms such as \texttt{doctor} and \texttt{patient}, indicating successful recovery of the semantic domain even if the exact specific token is not fully resolved.


\begin{table}[htbp]
  \centering
  \caption[Effect of target centering on recovery]{Recovery with and without target centering, over the full vocabulary of 128\,256 tokens. Profiled words are evaluated on held-out traces of the training vocabulary; unseen words are 25 words excluded from training entirely.}
  \label{tab:centering}
  \begin{tabular}{llcc}
  \toprule
    Setting & Metric & Baseline & Centering \\
    \midrule
    Unseen words   & Median rank        & 1\,327 & 443 \\
                   & Top-100            & 11.8\% & 28.7\% \\
    \bottomrule
  \end{tabular}
\end{table}

\vspace{0.5ex}\noindent\textbf{Optimizing for Open-Vocabulary Recovery.}
The preceding loss ablation evaluates recovery among the profiled targets. In contrast, Table~\ref{tab:loss-vocab} evaluates unseen-word recovery over the full vocabulary of 128\,256 tokens. We explore multiple options to guide the model's learning for this task, as the previous objectives compare predictions only against words seen during training, whereas the attack ranks predictions against the entire 128\,256-token vocabulary.
To align the training process with the attack's end goal, our generative model is optimized using full-vocabulary cross-entropy augmented with a hardest-negative margin. This objective directly matches the evaluation pipeline: the predicted hidden state is multiplied by the LLM's output head to produce a score for every token, and cross-entropy against the true token encourages a high ranking across the full 128\,256-token vocabulary. The hardest-negative margin additionally pushes the prediction away from the highest-scoring incorrect class. For comparison, we evaluate three baseline objectives: a standard cosine similarity loss (which optimizes only against profiled words), cosine similarity with a hardest-negative margin, and full-vocabulary cross-entropy with an additive angular margin~\cite{deng2019arcface}. As shown in Table~\ref{tab:loss-vocab}, shifting from cosine similarity to full-vocabulary cross-entropy drastically improves generalization to unseen words, reducing the median rank from 443 to 55 and more than doubling the Top-100 recovery rate (from 28.7\% to 63.8\%). The addition of the hardest-negative margin further refines this, achieving the highest Top-100 recovery at 67.0\% and a median rank of 54.

\begin{table}[htbp]
  \centering
  \caption[Training objective and unseen-word recovery]{Unseen-word recovery by training objective, over the full vocabulary of 128\,256 tokens. Top-100 is the fraction of traces for which the correct token falls within the 100 highest-ranked tokens.}
  \label{tab:loss-vocab}
  \small
  \resizebox{\linewidth}{!}{
  \begin{tabular}{@{}lrrr@{}}
    \toprule
    Training objective & Top-5 \% & Top-100 \% & Median rank \\
    \midrule
    Cosine                                     & 2.6 & 28.7 & 443 \\
    Hardest-negative margin                    & \textbf{7.3} & 28.2 & 365 \\
    Full-vocabulary cross-entropy              & 0.0 & 63.8 & 55 \\
    \quad + angular margin                     & 0.0 & 63.5 & \textbf{53} \\
    \quad + hardest-negative margin            & 0.0 & \textbf{67.0} & 54 \\
    \bottomrule
  \end{tabular}
  }
\end{table}
\label{sec:inversion}
\subsection{Semantic Prompt Recovery}
The preceding sections establish the recovery of a single emitted token. We now examine whether the EM-predicted state retains sufficient information to semantically recover the full user prompt using the inversion model described in Section~\ref{sec:method-prompt-inversion}. During evaluation, the recovery relies exclusively on the raw EM-predicted hidden state.

The end-to-end attack successfully infers the user prompt from an electromagnetic measurement, producing text that captures its semantic content. Table~\ref{tab:inv-examples} shows representative inferred prompts directly from EM-predicted states. 

\begin{table}[htbp]
  \centering
  \caption[Semantic prompt recovery examples]{Prompts recovered from EM traces. Every submitted prompt had the form \texttt{Say the words 'hello WORD' and nothing else, with no punctuation.}, a phrasing the recovery model was never trained on. The upper block shows exact semantic recoveries, and the lower block illustrates cases with partial recovery.}
  \label{tab:inv-examples}
  \small
  \begin{tabular}{@{}llp{4.5cm}@{}}
    \toprule
    Submitted word & Recovered & Recovered prompt \\
    \midrule
    coping       & yes & \texttt{Write 'hello coping' once, with no extra words.} \\
    pregnancy    & yes & \texttt{Please echo back the phrase 'hello pregnancy' verbatim.} \\
    \midrule
    influenza    & no  & \texttt{Write 'hello influen' once, with no extra words.} \\
    sadness      & no  & \texttt{Write 'hello sleep' once, with no extra words.} \\
    \bottomrule
  \end{tabular}
\end{table}
Since the capture templates were deliberately excluded from the training data, the model relies on semantically recovering the instruction in its own words rather than simply memorizing the original phrasing. The results indicate that the attack successfully recovers the semantic intent of the query, achieving a mean cosine similarity of $0.82$ with the original private user prompt across a test set of $1\,024$ evaluated prompts. Furthermore, the model extracts the exact embedded secret keyword in $88.7\%$ of these traces.


For the remaining 11.3\% of cases that result in a mismatch error, the target word is frequently truncated (e.g., \emph{influen}). These errors appear to cluster near the correct answer in the semantic space. This proximity confirms that the predicted hidden state successfully captures the broad semantic content of the target token, even when it lacks the precision to resolve the exact word.


\section{Discussion}
\label{ch:discussion}

\vspace{0.5ex}\noindent\textbf{Origins of the Leakage.}
Our empirical analysis of the \texttt{llama.cpp} execution flow discovered the temporal location of the leakage identified by the spatial masking and Grad-CAM evaluations. The initial segment of the execution window consists of CPU scheduling overhead and loading the required model weights. The most discriminative leakage resides in the middle of the window, which directly corresponds to the core matrix-vector dot products of the LM Head projection. Finally, the latter portion of the trace corresponds to de-quantization, which carries significantly less token-specific information.

\vspace{0.5ex}\noindent\textbf{Limitations.}
In a real-world deployment, an attacker would face cross-session variations and dynamically shifting CPU frequencies. However, these limitations can be addressed. Cross-session variations can be mitigated by capturing profiling traces across multiple sessions and operating conditions. Automated pattern-matching techniques (such as cross-correlation with known trace templates) can detect and align the sequential decode phases without hardware triggers. Variations in CPU frequency can be addressed by utilizing more specialized measurement equipment.

Our study focuses on CPU-based LLM inference, but many real-world edge deployments rely on GPUs,
which may pose challenges for the attacker due to higher clock frequencies and parallelization.
Future work may investigate adapting our methods to GPU-based deployments, for example,
by exploiting GPU-specific execution and emanation patterns.

\vspace{0.5ex}\noindent\textbf{Countermeasures.}
Physical side-channel leakage during edge LLM inference may be mitigated by interventions at different layers of the system. Hardware-level shielding and physical enclosures can degrade the signal-to-noise ratio of EM emanations. At the software and microarchitectural levels, execution randomization (e.g., randomizing the sequence of dot products) breaks the temporal alignment necessary for profiling attacks. Furthermore, algorithmic masking techniques~\cite{10.1007/978-3-642-38348-9_9}, traditionally used in cryptography, could be adapted to obfuscate the hidden state vectors and vocabulary matrix during the LM Head computation, decoupling the intermediate physical states from the true token values. We anticipate that developing low-overhead defenses against side-channel leakage in edge LLM inference will be a fruitful direction for future research.

\section{Conclusion}
\label{ch:conclusion}

We demonstrate that executing LLMs on edge devices can expose generated tokens and user inputs through physical side channels. By analyzing \texttt{llama.cpp}, we identify the LM~Head matrix multiplication as a source of electromagnetic token leakage, which standard convolutional networks can effectively exploit when converted to time-frequency spectrograms. We present three progressive attacks: 1) a closed-set classifier that identifies profiled keywords with 97.2\% accuracy; 2) an open-vocabulary generative model that reconstructs the internal hidden state to decode unseen tokens across the whole 128\,256-token vocabulary, achieving 67.0\% top-100 recovery with a median rank of 54; 3) a prompt-recovery attack that leverages the recovered hidden LLM states to semantically reconstruct the user's original query, which extracts the embedded secret keyword in 88.7\% of the cases.

This work establishes a foundation for the physical side-channel analysis of LLMs, demonstrating that local execution alone does not guarantee privacy. Our findings provide a basis for developers and hardware manufacturers to mitigate physical leakage through 
countermeasures such as EM shielding, LM Head masking, and execution randomization.
In conclusion, given the surge in LLM deployment, we argue that physical side-channel attacks against LLM inference and corresponding defenses warrant greater attention from both researchers and practitioners.

\section*{Ethical Considerations}
\label{sec:ethics}
Exposing this attack surface contributes to security awareness by making an invisible leakage channel measurable and reproducible. The objective of this work is to establish the vulnerability and characterize its practical limitations, providing a foundation for developers and hardware manufacturers to account for physical leakage and evaluate appropriate countermeasures for on-device LLMs.

\section*{Open Science}
\label{sec:openscience}
The source code for the attack framework, sample EM traces, trained models, and hidden state representation samples will be made publicly available upon publication.

\section*{LLM Usage Considerations}
\label{sec:llm_usage}
LLMs were utilized to refine grammar and stylistic flow, and to assist with code debugging.

\bibliographystyle{IEEEtran}
\bibliography{refs}

\begin{thebibliography}{10}
\providecommand{\url}[1]{#1}
\csname url@samestyle\endcsname
\providecommand{\newblock}{\relax}
\providecommand{\bibinfo}[2]{#2}
\providecommand{\BIBentrySTDinterwordspacing}{\spaceskip=0pt\relax}
\providecommand{\BIBentryALTinterwordstretchfactor}{4}
\providecommand{\BIBentryALTinterwordspacing}{\spaceskip=\fontdimen2\font plus
\BIBentryALTinterwordstretchfactor\fontdimen3\font minus
  \fontdimen4\font\relax}
\providecommand{\BIBforeignlanguage}[2]{{%
\expandafter\ifx\csname l@#1\endcsname\relax
\typeout{** WARNING: IEEEtran.bst: No hyphenation pattern has been}%
\typeout{** loaded for the language `#1'. Using the pattern for}%
\typeout{** the default language instead.}%
\else
\language=\csname l@#1\endcsname
\fi
#2}}
\providecommand{\BIBdecl}{\relax}
\BIBdecl

\bibitem{gerganov2023llamacpp}
\BIBentryALTinterwordspacing
G.~Gerganov and contributors, ``llama.cpp,'' GitHub repository, 2026. [Online].
  Available: \url{https://github.com/ggml-org/llama.cpp}
\BIBentrySTDinterwordspacing

\bibitem{shen2025agilequantactivationguidedquantizationfaster}
\BIBentryALTinterwordspacing
X.~Shen, P.~Dong, L.~Lu, Z.~Kong, Z.~Li, M.~Lin, C.~Wu, and Y.~Wang,
  ``Agile-quant: Activation-guided quantization for faster inference of llms on
  the edge,'' 2025. [Online]. Available: \url{https://arxiv.org/abs/2312.05693}
\BIBentrySTDinterwordspacing

\bibitem{raspberrypi5}
\BIBentryALTinterwordspacing
{Raspberry Pi Ltd}, ``Raspberry pi 5 product brief,'' 2023, accessed 2025.
  [Online]. Available:
  \url{https://www.raspberrypi.com/products/raspberry-pi-5/}
\BIBentrySTDinterwordspacing

\bibitem{nvidia_jetson_orin}
\BIBentryALTinterwordspacing
{NVIDIA Corporation}, ``{NVIDIA} {Jetson} {Orin} {Nano} system-on-module,''
  Product page,
  \url{https://www.nvidia.com/en-us/autonomous-machines/embedded-systems/jetson-orin/},
  2023. [Online]. Available:
  \url{https://www.nvidia.com/en-us/autonomous-machines/embedded-systems/jetson-orin/}
\BIBentrySTDinterwordspacing

\bibitem{hailo8l_aikit}
\BIBentryALTinterwordspacing
{Hailo Technologies}, ``Hailo-8{L} {AI} accelerator and {Raspberry} {Pi} {AI}
  {Kit},'' Product page, \url{https://hailo.ai/products/hailo-8l/}, 2024.
  [Online]. Available: \url{https://hailo.ai/products/hailo-8l/}
\BIBentrySTDinterwordspacing

\bibitem{prajapati2024medaide}
\BIBentryALTinterwordspacing
A.~Basit, K.~Hussain, M.~A. Hanif, and M.~Shafique, ``Medaide: Leveraging large
  language models for on-premise medical assistance on edge devices,'' 2024.
  [Online]. Available: \url{https://arxiv.org/abs/2403.00830}
\BIBentrySTDinterwordspacing

\bibitem{su2024physiollm}
\BIBentryALTinterwordspacing
C.~M. Fang, V.~Danry, N.~Whitmore, A.~Bao, A.~Hutchison, C.~Pierce, and
  P.~Maes, ``Physiollm: Supporting personalized health insights with wearables
  and large language models,'' 2024. [Online]. Available:
  \url{https://arxiv.org/abs/2406.19283}
\BIBentrySTDinterwordspacing

\bibitem{zhong2025considerations}
\BIBentryALTinterwordspacing
X.~Zhong, S.~Li, Z.~Chen, L.~Ge, D.~Yu, S.~Wang, L.~You, and H.~Shang,
  ``Considerations for patient privacy of large language models in health care:
  Scoping review,'' \emph{J Med Internet Res}, vol.~27, p. e76571, Nov 2025.
  [Online]. Available: \url{https://www.jmir.org/2025/1/e76571}
\BIBentrySTDinterwordspacing

\bibitem{demir2025legalguardian}
\BIBentryALTinterwordspacing
M.~M. Demir, H.~T. Otal, and M.~A. Canbaz, ``Legalguardian: A
  privacy-preserving framework for secure integration of large language models
  in legal practice,'' 2025. [Online]. Available:
  \url{https://arxiv.org/abs/2501.10915}
\BIBentrySTDinterwordspacing

\bibitem{xue2026mindchat}
\BIBentryALTinterwordspacing
D.~Xue, J.~Tu, M.~Wang, X.~Yan, F.~Liu, and J.~Hu, ``Towards privacy-preserving
  mental health support with large language models,'' 2026. [Online].
  Available: \url{https://arxiv.org/abs/2601.01993}
\BIBentrySTDinterwordspacing

\bibitem{mandal2025towards}
A.~Mandal, T.~Chakraborty, and I.~Gurevych, ``Towards privacy-aware mental
  health ai models,'' \emph{Nature Computational Science}, vol.~5, no.~10, pp.
  863--874, 2025.

\bibitem{agrawal2002em}
D.~Agrawal, B.~Archambeault, J.~R. Rao, and P.~Rohatgi, ``The em
  side---channel(s),'' in \emph{Cryptographic Hardware and Embedded Systems -
  CHES 2002}, B.~S. Kaliski, {\c{c}}.~K. Ko{\c{c}}, and C.~Paar, Eds.\hskip 1em
  plus 0.5em minus 0.4em\relax Berlin, Heidelberg: Springer Berlin Heidelberg,
  2003, pp. 29--45.

\bibitem{PEETERS200752}
\BIBentryALTinterwordspacing
E.~Peeters, F.-X. Standaert, and J.-J. Quisquater, ``Power and electromagnetic
  analysis: Improved model, consequences and comparisons,'' \emph{Integration},
  vol.~40, no.~1, pp. 52--60, 2007, embedded Cryptographic Hardware. [Online].
  Available:
  \url{https://www.sciencedirect.com/science/article/pii/S0167926005000647}
\BIBentrySTDinterwordspacing

\bibitem{10.1007/3-540-45418-7_17}
J.-J. Quisquater and D.~Samyde, ``Electromagnetic analysis (ema): Measures and
  counter-measures for smart cards,'' in \emph{Smart Card Programming and
  Security}, I.~Attali and T.~Jensen, Eds.\hskip 1em plus 0.5em minus
  0.4em\relax Berlin, Heidelberg: Springer Berlin Heidelberg, 2001, pp.
  200--210.

\bibitem{10.1007/3-540-44709-1_21}
K.~Gandolfi, C.~Mourtel, and F.~Olivier, ``Electromagnetic analysis: Concrete
  results,'' in \emph{Cryptographic Hardware and Embedded Systems --- CHES
  2001}, {\c{C}}.~K. Ko{\c{c}}, D.~Naccache, and C.~Paar, Eds.\hskip 1em plus
  0.5em minus 0.4em\relax Berlin, Heidelberg: Springer Berlin Heidelberg, 2001,
  pp. 251--261.

\bibitem{9586131}
E.~Karabulut and A.~Aysu, ``Falcon down: Breaking falcon post-quantum signature
  scheme through side-channel attacks,'' in \emph{2021 58th ACM/IEEE Design
  Automation Conference (DAC)}, 2021, pp. 691--696.

\bibitem{wang2024amplitude}
H.~Wang, ``Amplitude-modulated em side-channel attack on provably secure masked
  aes,'' \emph{Journal of Cryptographic Engineering}, vol.~14, no.~3, pp.
  537--549, 2024.

\bibitem{batina2018csinn}
\BIBentryALTinterwordspacing
L.~Batina, S.~Bhasin, D.~Jap, and S.~Picek, ``{CSI} {NN}: Reverse engineering
  of neural network architectures through electromagnetic side channel,'' in
  \emph{28th USENIX Security Symposium (USENIX Security 19)}.\hskip 1em plus
  0.5em minus 0.4em\relax Santa Clara, CA: USENIX Association, Aug. 2019, pp.
  515--532. [Online]. Available:
  \url{https://www.usenix.org/conference/usenixsecurity19/presentation/batina}
\BIBentrySTDinterwordspacing

\bibitem{9615240}
------, ``Sca strikes back: Reverse-engineering neural network architectures
  using side channels,'' \emph{IEEE Design \& Test}, vol.~39, no.~4, pp. 7--14,
  2022.

\bibitem{horvath2025barracuda}
\BIBentryALTinterwordspacing
P.~Horvath, L.~Chmielewski, L.~Weissbart, L.~Batina, and Y.~Yarom,
  ``{BarraCUDA}: Edge {GPUs} do leak {DNN} weights,'' in \emph{34th USENIX
  Security Symposium (USENIX Security 25)}.\hskip 1em plus 0.5em minus
  0.4em\relax Seattle, WA: USENIX Association, Aug. 2025, pp. 4017--4034.
  [Online]. Available:
  \url{https://www.usenix.org/conference/usenixsecurity25/presentation/horvath}
\BIBentrySTDinterwordspacing

\bibitem{horvath2025kraken}
\BIBentryALTinterwordspacing
P.~Horvath, I.~Shumailov, L.~Chmielewski, L.~Batina, and Y.~Yarom, ``Kraken:
  Higher-order em side-channel attacks on dnns in near and far field,'' 2026.
  [Online]. Available: \url{https://arxiv.org/abs/2603.02891}
\BIBentrySTDinterwordspacing

\bibitem{liu2025scaar}
\BIBentryALTinterwordspacing
Z.~Liu, S.~van Hoek, P.~Horváth, D.~Lauret, X.~Xu, and L.~Batina, ``Real-world
  edge neural network implementations leak private interactions through
  physical side channel,'' 2025. [Online]. Available:
  \url{https://arxiv.org/abs/2501.14512}
\BIBentrySTDinterwordspacing

\bibitem{dubey2024llama3}
\BIBentryALTinterwordspacing
A.~Grattafiori, A.~Dubey, A.~Jauhri, A.~Pandey, A.~Kadian, A.~Al-Dahle,
  A.~Letman, A.~Mathur, A.~Schelten, A.~Vaughan \emph{et~al.}, ``The llama 3
  herd of models,'' 2024. [Online]. Available:
  \url{https://arxiv.org/abs/2407.21783}
\BIBentrySTDinterwordspacing

\bibitem{horvath2024sok}
P.~Horv\'{a}th, D.~Lauret, Z.~Liu, and L.~Batina, ``Sok: neural network
  extraction through physical side channels,'' in \emph{Proceedings of the 33rd
  USENIX Security Symposium}, ser. SEC '24.\hskip 1em plus 0.5em minus
  0.4em\relax USA: USENIX Association, 2024.

\bibitem{wang2024imc}
\BIBentryALTinterwordspacing
Z.~Wang, F.-H. Meng, Y.~Park, J.~K. Eshraghian, and W.~D. Lu, ``Side-channel
  attack analysis on in-memory computing architectures,'' \emph{IEEE
  Transactions on Emerging Topics in Computing}, vol.~12, no.~1, p. 109–121,
  Jan. 2024. [Online]. Available:
  \url{http://dx.doi.org/10.1109/TETC.2023.3257684}
\BIBentrySTDinterwordspacing

\bibitem{gao2025iknow}
\BIBentryALTinterwordspacing
Z.~Gao, J.~Hu, F.~Guo, Y.~Zhang, Y.~Han, S.~Liu, H.~Li, and Z.~Lv, ``I know
  what you said: Unveiling hardware cache {Side-Channels} in local large
  language model inference,'' in \emph{34th USENIX Security Symposium (USENIX
  Security 25)}.\hskip 1em plus 0.5em minus 0.4em\relax Seattle, WA: USENIX
  Association, Aug. 2025, pp. 1649--1668. [Online]. Available:
  \url{https://www.usenix.org/conference/usenixsecurity25/presentation/gao-zibo}
\BIBentrySTDinterwordspacing

\bibitem{adiletta2025spill}
\BIBentryALTinterwordspacing
A.~Adiletta and B.~Sunar, ``Spill the beans: Exploiting cpu cache side-channels
  to leak tokens from large language models,'' 2025. [Online]. Available:
  \url{https://arxiv.org/abs/2505.00817}
\BIBentrySTDinterwordspacing

\bibitem{mcdonald2025whisper}
\BIBentryALTinterwordspacing
G.~McDonald and J.~B. Or, ``Whisper leak: a side-channel attack on large
  language models,'' 2025. [Online]. Available:
  \url{https://arxiv.org/abs/2511.03675}
\BIBentrySTDinterwordspacing

\bibitem{morris2023vec2text}
\BIBentryALTinterwordspacing
J.~X. Morris, V.~Kuleshov, V.~Shmatikov, and A.~M. Rush, ``Text embeddings
  reveal (almost) as much as text,'' 2023. [Online]. Available:
  \url{https://arxiv.org/abs/2310.06816}
\BIBentrySTDinterwordspacing

\bibitem{morris2024languagemodelinversion}
\BIBentryALTinterwordspacing
J.~X. Morris, W.~Zhao, J.~T. Chiu, V.~Shmatikov, and A.~M. Rush, ``Language
  model inversion,'' 2023. [Online]. Available:
  \url{https://arxiv.org/abs/2311.13647}
\BIBentrySTDinterwordspacing

\bibitem{10.1145/3786170.3788387}
\BIBentryALTinterwordspacing
Y.~Kong, S.~Bartolucci, F.~Caccioli, and G.~Destefanis, ``Separation of
  concerns for privacy-preserving llm adoption: A banking architecture
  framework,'' in \emph{Proceedings of the 3rd IEEE/ACM Workshop on Software
  Engineering Challenges in Financial Firms}, ser. FinanSE '26.\hskip 1em plus
  0.5em minus 0.4em\relax New York, NY, USA: Association for Computing
  Machinery, 2026, pp. 1--8. [Online]. Available:
  \url{https://doi.org/10.1145/3786170.3788387}
\BIBentrySTDinterwordspacing

\bibitem{chari2002template}
S.~Chari, J.~R. Rao, and P.~Rohatgi, ``Template attacks,'' in
  \emph{Cryptographic Hardware and Embedded Systems - CHES 2002}, B.~S.
  Kaliski, {\c{c}}.~K. Ko{\c{c}}, and C.~Paar, Eds.\hskip 1em plus 0.5em minus
  0.4em\relax Berlin, Heidelberg: Springer Berlin Heidelberg, 2003, pp. 13--28.

\bibitem{arm_neon_optimization}
\BIBentryALTinterwordspacing
{Arm}, ``Optimize the dot-product application with arm neon intrinsics.''
  [Online]. Available:
  \url{https://learn.arm.com/learning-paths/servers-and-cloud-computing/performix-get-started/neon-optimization/}
\BIBentrySTDinterwordspacing

\bibitem{lecroy_waverunner8404m}
\BIBentryALTinterwordspacing
{Teledyne LeCroy}, ``{WaveRunner 8000HD} high definition oscilloscopes
  datasheet,'' 2022, accessed 2025. [Online]. Available:
  \url{https://www.teledynelecroy.com/oscilloscope/waverunner-8000hd}
\BIBentrySTDinterwordspacing

\bibitem{grossmann1984decomposition}
A.~Grossmann and J.~Morlet, ``Decomposition of hardy functions into square
  integrable wavelets of constant shape,'' \emph{SIAM journal on mathematical
  analysis}, vol.~15, no.~4, pp. 723--736, 1984.

\bibitem{sao2024joint}
\BIBentryALTinterwordspacing
K.~Sao and M.~V. Chilukuri, ``Joint time-frequency analysis of time-varying
  signals for condition monitoring,'' \emph{AIP Conference Proceedings}, vol.
  2966, no.~1, p. 030015, 03 2024. [Online]. Available:
  \url{https://doi.org/10.1063/5.0189830}
\BIBentrySTDinterwordspacing

\bibitem{raihan2005wavelet}
S.~Raihan, Y.~Wen, and B.~Zeng, ``Wavelet: A new tool for business cycle
  analysis,'' \emph{Federal Reserve Bank of St. Louis Working Paper Series},
  vol.~50, 02 2005.

\bibitem{raffel2020t5}
\BIBentryALTinterwordspacing
C.~Raffel, N.~Shazeer, A.~Roberts, K.~Lee, S.~Narang, M.~Matena, Y.~Zhou,
  W.~Li, and P.~J. Liu, ``Exploring the limits of transfer learning with a
  unified text-to-text transformer,'' 2023. [Online]. Available:
  \url{https://arxiv.org/abs/1910.10683}
\BIBentrySTDinterwordspacing

\bibitem{deng2009imagenet}
J.~Deng, W.~Dong, R.~Socher, L.-J. Li, K.~Li, and L.~Fei-Fei, ``Imagenet: A
  large-scale hierarchical image database,'' in \emph{2009 IEEE Conference on
  Computer Vision and Pattern Recognition}, 2009, pp. 248--255.

\bibitem{he2016resnet}
K.~He, X.~Zhang, S.~Ren, and J.~Sun, ``Deep residual learning for image
  recognition,'' in \emph{{IEEE} Conference on Computer Vision and Pattern
  Recognition ({CVPR})}.\hskip 1em plus 0.5em minus 0.4em\relax {IEEE}, 2016,
  pp. 770--778.

\bibitem{tan2020efficientnetrethinkingmodelscaling}
\BIBentryALTinterwordspacing
M.~Tan and Q.~V. Le, ``Efficientnet: Rethinking model scaling for convolutional
  neural networks,'' 2020. [Online]. Available:
  \url{https://arxiv.org/abs/1905.11946}
\BIBentrySTDinterwordspacing

\bibitem{szegedy2016rethinking}
\BIBentryALTinterwordspacing
C.~Szegedy, V.~Vanhoucke, S.~Ioffe, J.~Shlens, and Z.~Wojna, ``Rethinking the
  inception architecture for computer vision,'' 2015. [Online]. Available:
  \url{https://arxiv.org/abs/1512.00567}
\BIBentrySTDinterwordspacing

\bibitem{kingma2015adam}
\BIBentryALTinterwordspacing
D.~P. Kingma and J.~Ba, ``Adam: A method for stochastic optimization,'' in
  \emph{International Conference on Learning Representations ({ICLR})}, 2015.
  [Online]. Available: \url{https://arxiv.org/abs/1412.6980}
\BIBentrySTDinterwordspacing

\bibitem{Selvaraju_2019}
\BIBentryALTinterwordspacing
R.~R. Selvaraju, M.~Cogswell, A.~Das, R.~Vedantam, D.~Parikh, and D.~Batra,
  ``Grad-cam: Visual explanations from deep networks via gradient-based
  localization,'' \emph{International Journal of Computer Vision}, vol. 128,
  no.~2, p. 336–359, Oct. 2019. [Online]. Available:
  \url{http://dx.doi.org/10.1007/s11263-019-01228-7}
\BIBentrySTDinterwordspacing

\bibitem{deng2019arcface}
\BIBentryALTinterwordspacing
J.~Deng, J.~Guo, J.~Yang, N.~Xue, I.~Kotsia, and S.~Zafeiriou, ``Arcface:
  Additive angular margin loss for deep face recognition,'' \emph{IEEE
  Transactions on Pattern Analysis and Machine Intelligence}, vol.~44, no.~10,
  pp. 5962--5979, Oct. 2022. [Online]. Available:
  \url{http://dx.doi.org/10.1109/TPAMI.2021.3087709}
\BIBentrySTDinterwordspacing

\bibitem{10.1007/978-3-642-38348-9_9}
E.~Prouff and M.~Rivain, ``Masking against side-channel attacks: A formal
  security proof,'' in \emph{Advances in Cryptology -- EUROCRYPT 2013},
  T.~Johansson and P.~Q. Nguyen, Eds.\hskip 1em plus 0.5em minus 0.4em\relax
  Berlin, Heidelberg: Springer Berlin Heidelberg, 2013, pp. 142--159.

\bibitem{zhao2024wildchat}
\BIBentryALTinterwordspacing
W.~Zhao, X.~Ren, J.~Hessel, C.~Cardie, Y.~Choi, and Y.~Deng, ``Wildchat: 1m
  chatgpt interaction logs in the wild,'' 2024. [Online]. Available:
  \url{https://arxiv.org/abs/2405.01470}
\BIBentrySTDinterwordspacing

\bibitem{nguyen2016msmarco}
\BIBentryALTinterwordspacing
P.~Bajaj, D.~Campos, N.~Craswell, L.~Deng, J.~Gao, X.~Liu, R.~Majumder,
  A.~McNamara, B.~Mitra, T.~Nguyen, M.~Rosenberg, X.~Song, A.~Stoica,
  S.~Tiwary, and T.~Wang, ``Ms marco: A human generated machine reading
  comprehension dataset,'' 2018. [Online]. Available:
  \url{https://arxiv.org/abs/1611.09268}
\BIBentrySTDinterwordspacing

\bibitem{wang2022supernatural}
\BIBentryALTinterwordspacing
Y.~Wang, S.~Mishra, P.~Alipoormolabashi, Y.~Kordi, A.~Mirzaei, A.~Arunkumar,
  A.~Ashok, A.~S. Dhanasekaran, A.~Naik, D.~Stap, E.~Pathak, G.~Karamanolakis,
  H.~G. Lai, I.~Purohit, I.~Mondal, J.~Anderson, K.~Kuznia, K.~Doshi, M.~Patel,
  K.~K. Pal, M.~Moradshahi, M.~Parmar, M.~Purohit, N.~Varshney, P.~R. Kaza,
  P.~Verma, R.~S. Puri, R.~Karia, S.~K. Sampat, S.~Doshi, S.~Mishra, S.~Reddy,
  S.~Patro, T.~Dixit, X.~Shen, C.~Baral, Y.~Choi, N.~A. Smith, H.~Hajishirzi,
  and D.~Khashabi, ``Super-naturalinstructions: Generalization via declarative
  instructions on 1600+ nlp tasks,'' 2022. [Online]. Available:
  \url{https://arxiv.org/abs/2204.07705}
\BIBentrySTDinterwordspacing

\bibitem{li2023chatdoctor}
\BIBentryALTinterwordspacing
Y.~Li, Z.~Li, K.~Zhang, R.~Dan, S.~Jiang, and Y.~Zhang, ``Chatdoctor: A medical
  chat model fine-tuned on a large language model meta-ai (llama) using medical
  domain knowledge,'' 2023. [Online]. Available:
  \url{https://arxiv.org/abs/2303.14070}
\BIBentrySTDinterwordspacing

\bibitem{rajpurkar2016squad}
\BIBentryALTinterwordspacing
P.~Rajpurkar, J.~Zhang, K.~Lopyrev, and P.~Liang, ``Squad: 100,000+ questions
  for machine comprehension of text,'' 2016. [Online]. Available:
  \url{https://arxiv.org/abs/1606.05250}
\BIBentrySTDinterwordspacing

\bibitem{joshi2017triviaqa}
\BIBentryALTinterwordspacing
M.~Joshi, E.~Choi, D.~S. Weld, and L.~Zettlemoyer, ``Triviaqa: A large scale
  distantly supervised challenge dataset for reading comprehension,'' 2017.
  [Online]. Available: \url{https://arxiv.org/abs/1705.03551}
\BIBentrySTDinterwordspacing

\bibitem{wei2024magicoder}
\BIBentryALTinterwordspacing
Y.~Wei, Z.~Wang, J.~Liu, Y.~Ding, and L.~Zhang, ``Magicoder: Empowering code
  generation with oss-instruct,'' 2024. [Online]. Available:
  \url{https://arxiv.org/abs/2312.02120}
\BIBentrySTDinterwordspacing

\bibitem{kopf2023openassistant}
\BIBentryALTinterwordspacing
A.~Köpf, Y.~Kilcher, D.~von Rütte, S.~Anagnostidis, Z.-R. Tam, K.~Stevens,
  A.~Barhoum, N.~M. Duc, O.~Stanley, R.~Nagyfi, S.~ES, S.~Suri, D.~Glushkov,
  A.~Dantuluri, A.~Maguire, C.~Schuhmann, H.~Nguyen, and A.~Mattick,
  ``Openassistant conversations -- democratizing large language model
  alignment,'' 2023. [Online]. Available:
  \url{https://arxiv.org/abs/2304.07327}
\BIBentrySTDinterwordspacing

\bibitem{chaudhary2023codealpaca}
S.~Chaudhary, ``Code alpaca: An instruction-following llama model for code
  generation,'' \url{https://github.com/sahil280114/codealpaca}, 2023.

\bibitem{conover2023dolly}
M.~Conover, M.~Hayes, A.~Mathur, J.~Xie, J.~Wan, S.~Shah, A.~Ghodsi,
  P.~Wendell, M.~Zaharia, and R.~Xin, ``Free dolly: Introducing the world's
  first truly open instruction-tuned {LLM},'' Databricks Blog, 2023,
  \url{https://www.databricks.com/blog/2023/04/12/dolly-first-open-commercially-viable-instruction-tuned-llm}.

\bibitem{welbl2017sciq}
\BIBentryALTinterwordspacing
J.~Welbl, N.~F. Liu, and M.~Gardner, ``Crowdsourcing multiple choice science
  questions,'' 2017. [Online]. Available:
  \url{https://arxiv.org/abs/1707.06209}
\BIBentrySTDinterwordspacing

\bibitem{han2023medalpaca}
\BIBentryALTinterwordspacing
T.~Han, L.~C. Adams, J.-M. Papaioannou, P.~Grundmann, T.~Oberhauser,
  A.~Figueroa, A.~Löser, D.~Truhn, and K.~K. Bressem, ``Medalpaca -- an
  open-source collection of medical conversational ai models and training
  data,'' 2025. [Online]. Available: \url{https://arxiv.org/abs/2304.08247}
\BIBentrySTDinterwordspacing

\bibitem{chalkidis2022lexglue}
\BIBentryALTinterwordspacing
I.~Chalkidis, A.~Jana, D.~Hartung, M.~Bommarito, I.~Androutsopoulos, D.~M.
  Katz, and N.~Aletras, ``Lexglue: A benchmark dataset for legal language
  understanding in english,'' 2022. [Online]. Available:
  \url{https://arxiv.org/abs/2110.00976}
\BIBentrySTDinterwordspacing

\bibitem{cobbe2021gsm8k}
\BIBentryALTinterwordspacing
K.~Cobbe, V.~Kosaraju, M.~Bavarian, M.~Chen, H.~Jun, L.~Kaiser, M.~Plappert,
  J.~Tworek, J.~Hilton, R.~Nakano, C.~Hesse, and J.~Schulman, ``Training
  verifiers to solve math word problems,'' 2021. [Online]. Available:
  \url{https://arxiv.org/abs/2110.14168}
\BIBentrySTDinterwordspacing

\end{thebibliography}
\newpage
\appendices
\section{Corpus Details}
\label{app:corpus-details}

\begin{table}[htbp]
  \centering
  \caption[Sources of the prompt recovery training corpus]{Composition of the prompt recovery corpus. Text from these sources was run through the victim model locally to generate corresponding hidden states. The victim's exact capture templates were excluded from training. In total, $2\,551\,308$ pairs are used for training, and the remainder are held out for evaluation.}
  \label{tab:corpus-sources}
  \scriptsize
  \setlength{\tabcolsep}{3pt}
  \renewcommand{\arraystretch}{0.85}
  \begin{tabular}{@{}lllrr@{}}
    \toprule
    Domain & Source corpus & & Prompts & Trained \\
    \midrule
    \multicolumn{5}{@{}l}{\textbf{Broad instructions}} \\
    \cmidrule(r){1-3}
    Mixed instructions & \texttt{one-million-instructions} & & 1\,200\,000 & 1\,175\,957 \\
    \midrule
    \multicolumn{5}{@{}l}{\textbf{Domain corpora}} \\
    \cmidrule(r){1-3}
    Conversation       & WildChat-1M                & \cite{zhao2024wildchat}      & 249\,904 & \\
    Web search         & MS MARCO                   & \cite{nguyen2016msmarco}     & 199\,998 & \\
    Privacy / PII      & \texttt{ai4privacy/pii-masking-300k} & & 177\,636 & \\
    Instructions       & Super-NaturalInstructions  & \cite{wang2022supernatural}  & 150\,000 & \\
    Medical            & ChatDoctor HealthCareMagic & \cite{li2023chatdoctor}      & 99\,999  & \\
    Question answering & SQuAD                     & \cite{rajpurkar2016squad}    & 87\,322  & \\
    Trivia QA           & TriviaQA                  & \cite{joshi2017triviaqa}     & 76\,490  & \\
    Code (issues)       & Magicoder OSS-Instruct    & \cite{wei2024magicoder}      & 74\,125  & \\
    Finance             & \texttt{gbharti/finance-alpaca} & & 58\,478 & \\
    Open chat           & OpenAssistant Conversations & \cite{kopf2023openassistant} & 30\,789 & \\
    Code (tasks)        & CodeAlpaca               & \cite{chaudhary2023codealpaca} & 19\,982 & \\
    General             & Databricks Dolly 15k     & \cite{conover2023dolly}      & 14\,741 & \\
    Science             & SciQ                     & \cite{welbl2017sciq}         & 11\,602 & \\
    Medical exams       & MedAlpaca / MedQA        & \cite{han2023medalpaca}      & 10\,163 & \\
    Legal               & LexGLUE (ECtHR)          & \cite{chalkidis2022lexglue} & 8\,972  & \\
    Mathematics         & GSM8K                    & \cite{cobbe2021gsm8k}        & 7\,473  & \\
    \cmidrule(r){1-3}
    \multicolumn{3}{@{}l}{\emph{subtotal, sixteen domain corpora}} & 1\,277\,674 & 1\,239\,351 \\
    \midrule
    \multicolumn{5}{@{}l}{\textbf{Generated templates}} \\
    \cmidrule(r){1-3}
    Slot-filling templates & generic phrasings & & 160\,000 & 136\,000 \\
    Victim templates       & the seventeen capture templates & & 51\,000 & 0 \\
    \midrule
    \multicolumn{3}{@{}l}{\textbf{Total}} & 2\,688\,674 & 2\,551\,308 \\
    \bottomrule
  \end{tabular}
\end{table}

\end{document}